\documentclass[twocolumn]{aastex62}
\usepackage[normalem]{ulem}
\useunder{\uline}{\ul}{}
\usepackage{hyperref}
\usepackage{lineno}

\usepackage{color}
\usepackage{amsmath,amssymb}

\hypersetup{linkcolor=cyan,citecolor=magenta,filecolor=yellow,urlcolor=blue}

\revised{\today}

\shorttitle{}
\shortauthors{}

\begin{document}

\title{X-ray and GeV afterglows and sub-TeV emission of GRB 180720B}

\author{R.~Moradi}
\affiliation{ICRA and Dipartimento di Fisica, Universit\`a  di Roma ``La Sapienza'', Piazzale Aldo Moro 5, I-00185 Roma, Italy}
\affiliation{ICRANet, Piazza della Repubblica 10, I-65122 Pescara, Italy}
\affiliation{INAF -- Osservatorio Astronomico d'Abruzzo,Via M. Maggini snc, I-64100, Teramo, Italy}

\author{Liang~Li}
\affiliation{ICRA and Dipartimento di Fisica, Universit\`a  di Roma ``La Sapienza'', Piazzale Aldo Moro 5, I-00185 Roma, Italy}
\affiliation{ICRANet, Piazza della Repubblica 10, I-65122 Pescara, Italy}

\author{J.~A.~Rueda}
\affiliation{ICRA and Dipartimento di Fisica, Universit\`a  di Roma ``La Sapienza'', Piazzale Aldo Moro 5, I-00185 Roma, Italy}
\affiliation{ICRANet, Piazza della Repubblica 10, I-65122 Pescara, Italy}
\affiliation{ICRANet-Ferrara, Dipartimento di Fisica e Scienze della Terra, Universit\`a degli Studi di Ferrara, Via Saragat 1, I--44122 Ferrara, Italy}
\affiliation{Dipartimento di Fisica e Scienze della Terra, Universit\`a degli Studi di Ferrara, Via Saragat 1, I--44122 Ferrara, Italy}
\affiliation{INAF, Istituto di Astrofisica e Planetologia Spaziali, Via Fosso del Cavaliere 100, 00133 Rome, Italy}

\author{R.~Ruffini}
\affiliation{ICRA and Dipartimento di Fisica, Universit\`a  di Roma ``La Sapienza'', Piazzale Aldo Moro 5, I-00185 Roma, Italy}
\affiliation{ICRANet, Piazza della Repubblica 10, I-65122 Pescara, Italy}
\affiliation{INAF,Viale del Parco Mellini 84, 00136 Rome, Italy}

\author{N.~Sahakyan}
\affiliation{ICRANet, Piazza della Repubblica 10, I-65122 Pescara, Italy}
\affiliation{ICRANet-Armenia, Marshall Baghramian Avenue 24a, Yerevan 0019, Republic of Armenia}

\author{Y.~Wang}
\affiliation{ICRA and Dipartimento di Fisica, Universit\`a  di Roma ``La Sapienza'', Piazzale Aldo Moro 5, I-00185 Roma, Italy}
\affiliation{ICRANet, Piazza della Repubblica 10, I-65122 Pescara, Italy}
\affiliation{INAF -- Osservatorio Astronomico d'Abruzzo,Via M. Maggini snc, I-64100, Teramo, Italy}
\email{rahim.moradi@inaf.it, liang.li@icranet.org, jorge.rueda@icra.it,\\ruffini@icra.it, narek.sahakyan@icranet.org, yu.wang@icranet.org}  


\begin{abstract}
 GRB 180720B, observed by {\it Fermi}-GBM, with redshift $z=0.653$, isotropic energy $E_{\rm iso}=5.92\times 10^{53}$~erg, and X-ray afterglow observed by the XRT onboard the \textit{Neil Gehrels Swift} satellite, is here classified as a Binary-driven Hypernova I (BdHN I). BdHN I are long GRBs with a binary progenitor composed of a carbon-oxygen core (CO$_{\rm core}$) and a neutron star (NS) companion with orbital period $\sim 5$~min. The gravitational collapse of the CO$_{\rm core}$ generates a supernova (SN) and a new NS ($\nu$NS) at its center. The SN hypercritical accretion onto the companion NS triggers its gravitational collapse forming a black hole (BH). An electrodynamical process near the BH horizon leads to the \textit{long-lasting} GeV emission with power-law luminosity $L_{\rm GeV}\propto t^{-1.19\pm0.04}$, powered by the BH rotational energy. Correspondingly, we determine the BH mass and spin. The $\nu$NS via its pulsar-like emission and fallback accretion injects energy into the magnetized SN ejecta generating synchrotron radiation. This explains the \textit{long-lasting} X-ray afterglow with power-law luminosity $L_X \propto t^{-1.48\pm 0.32}$, energized by the $\nu$NS rotational energy. We apply this to GRB 180720B determining the $\nu$NS magnetic field and spin. We also analyze the GRB 180720B emission observed by the High-Energy Stereoscopic System (H.E.S.S.), at $100$--$440$~GeV energies, $10.1$--$12.1$~h after the Fermi-GBM trigger. We propose that this \textit{short-term} radiation of energy $2.4\times 10^{50}$~erg and duration $\sim 10^3$~s, is powered by a ``\textit{glitch}'' event that suddenly injects relativistic electrons into the $\nu$NS magnetosphere during its slowing down phase.
\end{abstract}


\section{Introduction} \label{sec:intro}

On $20$ July $2018$, at 14:21:39.65 universal time (UT), the Fermi Gamma-Ray Burst Monitor (GBM) was triggered (trigger 553789304/180720598) and located GRB 180720B. The GBM lightcurve comprises a very bright pulse, with numerous overlapping pulses with a duration of $T_{90}=49$~s in the observer frame $50$--$300$~keV \citep{2018GCN.22981....1R}, classifying this burst as a long GRB.

This GRB triggered as well the Swift Burst Alert Telescope (BAT) (trigger 848890) $5$~s after the GBM trigger time (hereafter $T_0$). Its light-curve as Fermi-GBM showed a multi-peaked structure with a duration of $\sim 150$~s. 

The light-curve is also seen up to $\sim 15$~MeV as reported by the Konus-Wind team \citep{2018GCN.23011....1F}. It shows also a multi-peaked structure with a total duration of $T_{\rm 100}$ $\sim 125$~s.

Multiwavelength follow-up observations were performed up to $T_0 + 3\times 10^5$~s by VLT/X-shooter spectrograph which measured a redshift of $z = 0.653$ \citep{2018GCN.22996....1V}. This allows us to adopt in this article the cosmological rest-frame time, $t_{\rm rf}$, and evaluate the energy of GRB 180720B.  

\begin{figure}
\centering
\includegraphics[width=\hsize,clip]{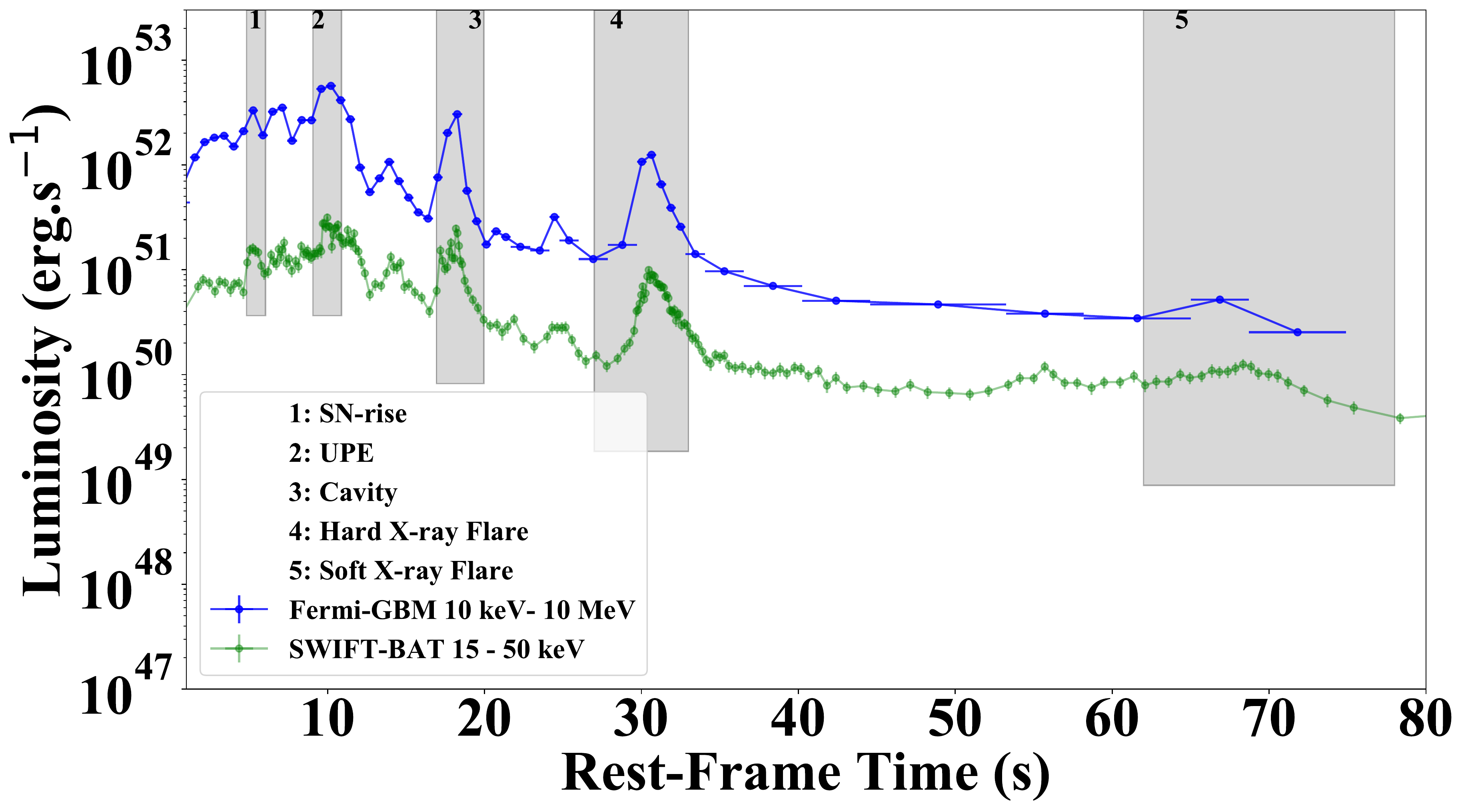}
\caption{The five episodes in GRB 180720B in the first $80$~s of the burst, in the cosmological rest-frame of the source, after the Fermi-GBM trigger time. The blue light-curve shows the luminosity deduced from Fermi-GBM in the energy range from $10$~keV to $10$~MeV. The green light-curve is the luminosity from Swift-BAT, in the $15-50$~keV energy range. The two satellites show consistent spiky structures and evolution.}
\label{fig:GBM_BAT_Linear}
\end{figure}

The above Swift-BAT, Fermi-GBM and Konus-Wind observations in keV-MeV range traditionally characterize the GRB prompt emission that includes five distinct episodes in the first $80$~s in GRB 180720B; see Fig.~\ref{fig:GBM_BAT_Linear}. 

\begin{figure*}
\centering
(a)\includegraphics[width=0.47\hsize,clip]{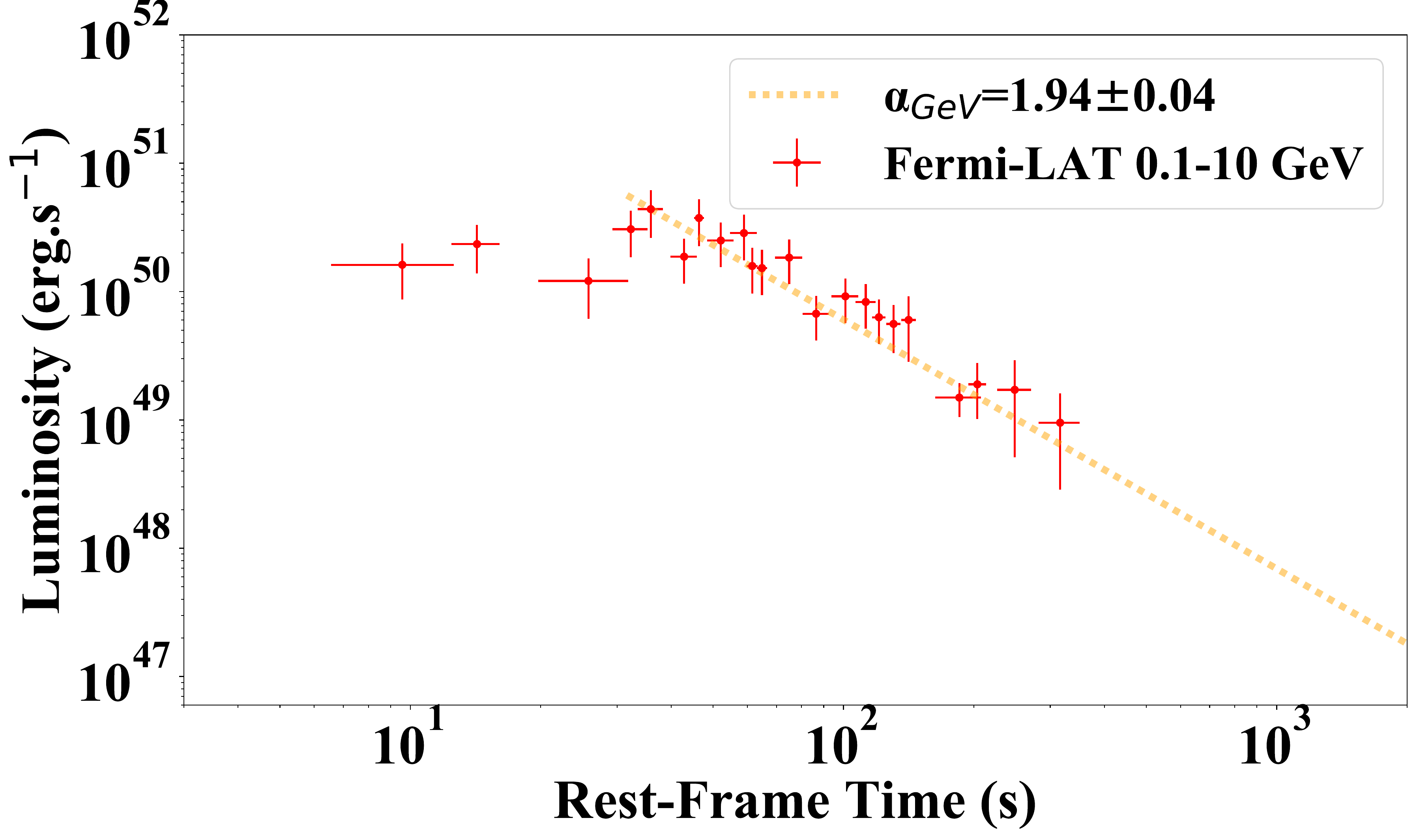}
(b)\includegraphics[width=0.47\hsize,clip]{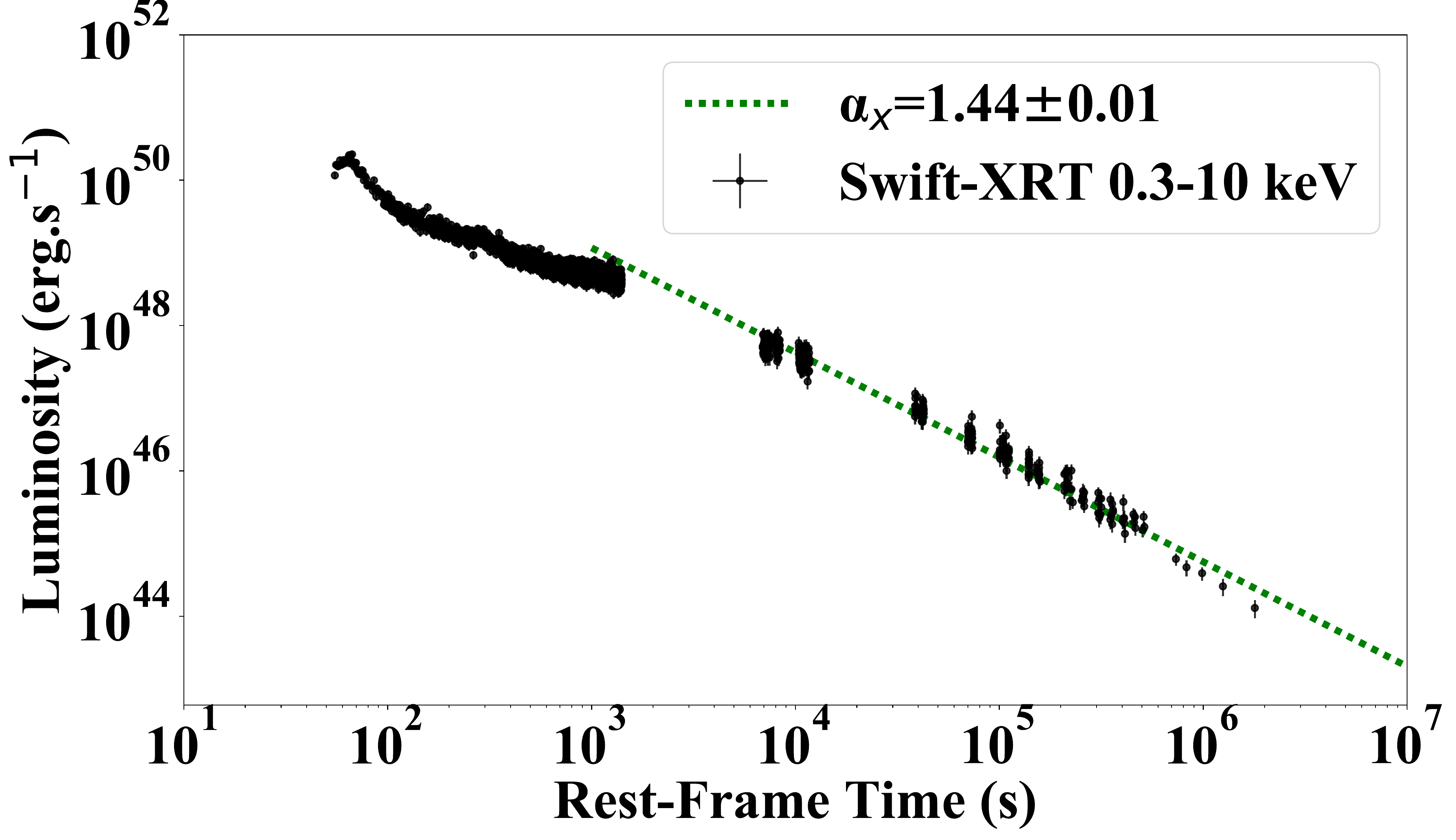}
(c)\includegraphics[width=0.47\hsize,clip]{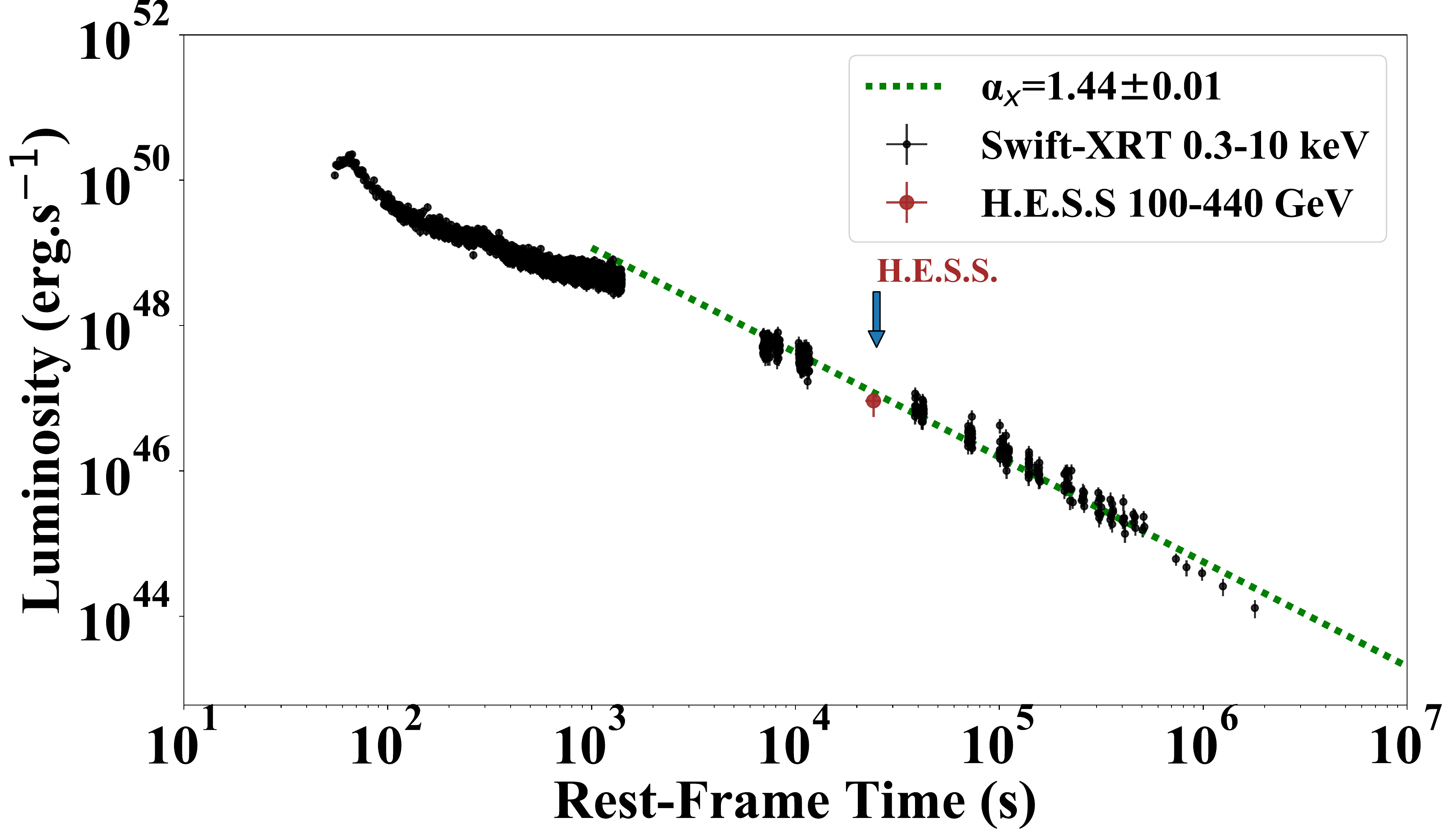}
\caption{\textbf{(a)}: The luminosity light-curve of GRB 180720B in $0.1$--$10$~GeV as observed by Fermi-LAT. The emission starts at $t_{\rm rf}=7.01$~s. The luminosity increases up to $t_{\rm rf} \sim 40$~s. After $t_{\rm rf} \sim 40$~s the GeV luminosity follows a temporal decaying luminosity of $L_{\rm GeV}= 4.6 \times 10^{53}~ t^{-1.94\pm0.0.13}$~erg~s$^{-1}$. \textbf{(b)}: The luminosity light-curve of GRB 180720B in $0.3$--$10$~keV as observed by Swift-XRT. The observation starts at $t_{\rm rf}=52$~s with a temporal decaying luminosity of $L_X =2.5 \times 10^{53} t^{-1.44\pm 0.01}$~erg~s$^{-1}$. \textbf{(c)}: The $100$--$440$~GeV emission as observed by H.E.S.S. The emission starts at $t_{\rm rf}=21956.5$~s and lasts for almost $1.2$~h. Its luminosity fits the luminosity of the X-ray in this time interval.}
\label{fig:GeV}
\end{figure*}

From $t_{\rm rf}=52$~s, the Swift-XRT started observation in the $0.3$--$10$~keV band and gave a clear evidence for the already existing X-ray afterglow \citep{2018GCN.22975....1S}. In addition to the above observations, GRB 180720B presents high energy radiations in the $0.1$--$10$~GeV detected by the Fermi-LAT instrument \citep{2018GCN.22980....1B}, and observations in the sub-TeV ($100$--$440$~GeV) band by the High Energy Stereoscopic System (H.E.S.S.) \citep{2019Natur.575..464A}. All this multiwavelength information makes this GRB almost unique and only comparable for the quality of data to GRB 190114C. In fact:
\begin{enumerate}
    \item 
    The GeV emission detected by Fermi-LAT starts at $t_{\rm rf}=7.01$~s with an increasing luminosity up to $t_{\rm rf}=40$~s, where it starts to decrease following the power-law $L_{\rm GeV}= 4.6 \times 10^{53}~ t^{-1.94\pm0.0.13}$~erg~s$^{-1}$; see Fig.~\ref{fig:GeV}(a).
    \item 
    The observation of the the X-ray afterglow luminosity observed by Swift-XRT starts at $t_{\rm rf}=52$~s with a temporal decaying luminosity of $L_X =2.5 \times 10^{53} t^{-1.44\pm 0.01}$~erg~s$^{-1}$; see Fig.~\ref{fig:GeV}(b).
    \item
    A clear sub--TeV short-duration emission was observed by H.E.S.S, starting at $t_{\rm rf}=21956.5$~s and lasting for nearly $1.2$~h, with a luminosity that perfectly fits the one of the X-ray in the same time interval; see Fig.~\ref{fig:GeV}(c).
\end{enumerate}

Due to its release of energy at a relatively nearby distance, the quantity and the quality of data are within the best observed ones in the GRB history. The flux intensity in the range of the  Fermi-GBM and Fermi-LAT makes this GRB as one of the best samples for spectral properties of the MeV-GeV radiation. Unlike GRB 130427A, where the Fermi-GBM data were affected by a pile-up that significantly deformed the spectrum (see details in \citealp{2014Sci...343...42A, 2015ApJ...798...10R}), this burst was not so intense to saturate the GBM detectors. 

For this reason, we proceed in the following section~\ref{sec:bdhn} to our analysis of this source by introducing a theoretical model with a series of Episodes each characterized by specific spectral features, to be uniquely identified by a time-resolved spectral analysis. We briefly introduce the Binary-driven Hypernova (BdHN) theory and its components and their predicted spectral properties; see e.g. \citet{2018ApJ...852...53R, 2018ApJ...869..151R}.

We present our time-resolved spectral analysis during the first $80$~s of the burst in order to identify the five different Episodes: 1) the rise of supernova (``\textit{SN-rise}''), 2) the ultra-relativistic prompt emission (UPE) phase, 3) the cavity, 4) the hard X-ray flare (HXF) and, finally, 5) the soft X-ray flare (SXF), predicted by the BdHN theory.
The data procedure of the time-resolved spectral analysis please refer to \cite{Li2019a, Li2019b, Li2020} and \cite{Li2019,Li2020b,Li2021} for details.

In section~\ref{sec:aftert90}, we proceed the data analysis of GRB 180720B following the $T_{\rm 90}$. We first present the X-ray afterglow originating from the SN hypercritical accretion onto the $\nu$NS as well as the GeV emission originating from the SN hypercritical accretion onto the BH. We then explicitly indicate the two different power-law luminosities characterising the X-ray and GeV emissions measured in the rest-frame of the source.

In section~\ref{sec:BH-NS-Hess}, using the properties of the ``\textit{inner engine}'' introduced in \citet{2019ApJ...886...82R, 2020EPJC...80..300R} for GRB 130427A, we here apply for GRB 180720B. We determine the magnetic field strength, the mass and the spin of the newly born BH. Similarly, we recall our results in \citet{2018ApJ...869..101R, 2019ApJ...874...39W, 2020ApJ...893..148R} showing that the X-ray afterglow radiation originates from the hypercritical accretion of the SN onto the $\nu$NS  we determine the period and the magnetic field structure of the $\nu$NS. Finally we summarize the general outlook of X-ray afterglow emission observed by Swift-XRT as well as the GeV emission observed by Fermi-LAT, and we propose, based on BdHN scenario, the connection of the H.E.S.S. radiation to the X-ray afterglow. 

In section~\ref{sec:glitch}, we recall the general formulation of the glitches in pulsars and propose to identify the energy release observed by H.E.S.S. to originate from the relativistic magnetohydrodynamic process occurring during the glitch event in the $\nu$NS. Finally, in section~\ref{sec:conclusion}, we present the conclusion of this article.

\section{BdHN and their spectral properties }\label{sec:bdhn}

Let us now turn to our theoretical model for long GRBs, the BdHN model, and give evidence that GRB 180720B classifies as a BdHN I. The traditional approach of GRBs \citep{1992MNRAS.258P..41R, 1993ApJ...405..273W} has tacitly assumed that all GRBs originate from a single Kerr BH. The first successful example of considering binary nature of progenitors was highlighted by the study of binary neutron star merging originating short GRBs \citep{1992ApJ...395L..83N}. 

The analysis of binary progenitors composed of different  combinations of NS, CO$_{\rm core}$, white dwarf (WD) and BH,  was advanced and all different nine classes of GRBs were introduced in \citet{2016ApJ...832..136R, 2019ApJ...874...39W}.

Special classes of long GRBs, with T$_{90}>2$~s, were introduced originating from a binary system composed of a CO$_{\rm core}$ merging with a binary NS companion  \citep{2015ApJ...798...10R,2015ApJ...812..100B,2016ApJ...833..107B,2016ApJ...832..136R,2019ApJ...874...39W,2019ApJ...886...82R}.

The first characteristic event in such binary system is represented by the gravitational collapse of the CO$_{\rm core}$ leading to the formation of a SN and, contextually from the collapse of its iron core, to the formation of a $\nu$NS. This is additional to the NS binary companion. The rise of the SN, referred hereafter to as \textit{SN-rise}, lasts a few seconds and is indicated by the observation of a  characteristic spectrum (we give details below). The hypercritical accretion of the SN ejecta occurs both on the $\nu$NS and on the binary companion NS (see e.g. Fig.~$5$ in \citealp{2019ApJ...871...14B}). The electrons which escape from the $\nu$NS magnetosphere are injected into the expanding magnetized HN ejecta giving origin to the X-ray afterglow via synchrotron radiation \citep{2018ApJ...869..101R}. The fit of the X-ray afterglow data with the synchrotron radiation in the HN ejecta shows that the magnetic field at $\sim 10^{12}$~cm is $B \sim 10^5$~G and decreases linearly with the radial distance. This behavior is indeed expected from the toroidal component of the $\nu$NS magnetic field at large distances well beyond the light cylinder \citep[see, e.g.,][for details]{2018ApJ...869..101R, 2019ApJ...874...39W, 2020ApJ...893..148R}. Therefore, this synchrotron emission of electrons occurs in the optically thin region of the HN ejecta, which expands at mildly-relativistic velocity, $v\approx 0.1c$, in the $\nu$NS magnetic field, at distances above $10^{12}$~cm \citep[see, e.g.,][]{2019ApJ...874...39W}.

Further signatures in the evolution occurs from the hypercritical accretion of the SN onto the companion NS and are a strong function of the binary period. For binary periods $\sim 5$~min, the hypercritical accretion of the SN ejecta onto the companion NS brings it to overcome the critical mass, with consequent gravitational collapse leading to the formation of a rotating BH \citep{2016ApJ...833..107B,2016ApJ...832..136R,2019ApJ...886...82R}. 

The formation of the BH is heralded by the onset of the emission of the GeV radiation; see Fig.~\ref{fig:GeV}. This GeV radiation originates from the \textit{inner engine} of the GRB originating from the interaction of a Kerr BH with a magnetic field aligned with the BH rotation axis, the Papapetrou-Wald solution; see \citep{2019ApJ...886...82R,2020EPJC...80..300R}.

Each Episode in a BdHN is characterised by precise spectral features, all visible thanks to the quality of the data of the Fermi and Swift satellites: 
\begin{enumerate}
    \item 
    The already mentioned \textit{SN-rise} characterized by the appearance of a blackbody component in the spectrum obtained from the Fermi-GBM; see  \citet{2012A&A...543A..10I,2014ApJ...793L..36F,2019arXiv190404162R,2019arXiv191012615L,2019ApJ...874...39W}. The spectrum of the \textit{SN-rise} of GRB 180720B is best fitted by a cut-off power-law plus a blackbody component (CPL+BB) model; see Fig.~\ref{fig:3episodes}(a).
   \item 
   The ultra-relativistic prompt emission (UPE) phase produced by the expansion and self-acceleration of an electron-positron-photon ($e^{\pm}\gamma$) plasma endowed with baryon load \citep{2007PhRvL..99l5003A,2009PhRvD..79d3008A}. As the plasma reaches transparency, a thermal radiation is emitted \citep{RSWX2,RSWX}, hence the UPE phase contains a specific spectra BB signature. The spectrum of the UPE phase of GRB 180720B is best fitted by a CPL+BB model; see Fig.~\ref{fig:3episodes} (b).
   \item 
   The ``\textit{cavity}'', first introduced for GRB 190114C in \citet{2019ApJ...883..191R}, is created by the gravitational collapse of the accreting companion NS originating the BH, and further depleted by the expanding $e^+~e^-~\gamma$ plasma.  Hydrodynamical numerical simulations presented in \citet{2019ApJ...883..191R} have shown that part of the $e^+~e^-~\gamma$ plasma is reflected off the walls of the \textit{cavity}. This reflected outflow and its observed properties coincide with the featureless emission observed in the Fermi-GMB data after the UPE phase of GRBs. The \textit{cavity} of GRB 180720B is best fitted by a single CPL; see Fig.~\ref{fig:3episodes} (c).
   \item 
   The hard X-ray Flares (HXF), marked by a PL+BB or a CPL spectrum, were first introduced in \citet{2018ApJ...869..151R} for GRB 151027A. They are produced by the acceleration process of the $e^+~e^-~\gamma$ plasma onto the SN eject, finally leading to the transition from an SN into a hypernova (HN). The HXF for GRB 180720B starts at $t_{\rm rf}\sim 30$~s, ands its spectrum is best fitted by a CPL model; see Fig.~\ref{fig:3episodes} (d).
   \item 
   The soft X-ray flares (SXF) are marked by a PL+BB spectrum identified in a sample of long GRBs by \citet{2018ApJ...852...53R} observed by Swift. They are produced by the acceleration process of the $e^+~e^-~\gamma$ plasma onto the SN eject, but with a different viewing angle with respect to the HXF \citep{2018ApJ...869..151R}, and occur at $t_{\rm rf}\sim 100$~s. The SXF for GRB 180720B starts at $t_{\rm rf}\sim 60$~s and its spectrum is best fitted by a PL+BB model; see Fig.~\ref{fig:3episodes} (e).   
\end{enumerate}
As pointed out by \citet{2018ApJ...869..151R} in the analysis of GRB 151027A, within BdHN model the UPE phase, the hard X-ray flare and the soft X-ray flare are not a causally connected sequence. They originate from the BH formation as observed from different viewing angles, as a consequence of the rotation period of $300$~s of the HN ejecta. We refer to section~\ref{sec:analysis180720B} for more details on the data analysis of five episodes in GRB 180720B.

From all the above, we infer that GRB 180720B can be considered a BdHN I \citep{2019ApJ...874...39W,2019ApJ...886...82R}. It is remarkable that we can assert that the necessary and sufficient condition to identify a long GRB as a BdHN I can be grounded on three characteristics (see  \citealp{2015ApJ...798...10R,2015ApJ...812..100B,2016ApJ...833..107B,2016ApJ...832..136R,2019ApJ...874...39W,2019ApJ...886...82R}, for more information):

\begin{enumerate}
    \item 
    A definite redshift. The redshift of GRB 180720B has been determined to be $z=0.653$, which allows us to express all our quantities in the rest-frame of the source.  
    \item 
    The isotropic gamma energy of BdHN I in the range of $10^{52} \lesssim E_{\rm iso}\lesssim 10^{54}$~erg and, with the peak energy in the range of $0.2$~MeV$<E_{\rm p,i}<2$~MeV \citep{2016ApJ...832..136R}. The isotropic energy for GRB 180720B is $E_{\rm iso}= 5.92 \times 10^{53}$~erg, determined by Fermi-GBM.
    \item 
    A \textit{long-lasting} X-ray afterglow with luminosity following a power-law decay with index of $\alpha_{X}= -1.48\pm 0.32$ and necessarily leads to an observable HN when the cosmological redshift is low enough. In the case of GRB 190114C, with $z= 0.42$, the SN has been indeed observed \citep{GCN23695,GCN23715,GCN23983}. 
    
\end{enumerate}

The reason of giving special attention to GRB 180720B is that this source, in addition of the observation of all the above five episodes, manifests (we refer to section~\ref{sec:aftert90} for more information about data analysis of GRB 180720B): 
\begin{itemize}
    \item 
    The unprecedented presence of a sub-TeV emission observed at $t_{\rm rf}\sim 6$~h by H.E.S.S. after the GBM trigger time  \citep{2019Natur.575..464A}. 
    \item 
    The presence of a GeV emission that starts at $t_{\rm rf}=7.01$~s; see Fig.~\ref{fig:GeV}(a), and is following a temporal decaying luminosity with a power-law index of $\alpha_{\rm GeV}=1.94\pm 0.13$.
    \item 
    The presence of an X-ray emission that at late time follows a temporal decaying luminosity with a power-law index of $\alpha_X=1.44\pm 0.01$.
\end{itemize}

The main issue is to understand if the sub-TeV emission observed by H.E.S.S. has the same nature of the prolonged X-ray emission originated from $\nu$NS or the prolonged GeV emission originated from the BH. 

\section{Time-resolved spectral analysis in the first $80$ seconds: the five episodes}\label{sec:analysis180720B}

\begin{figure*}
\centering
A\includegraphics[width=0.471\hsize,clip]{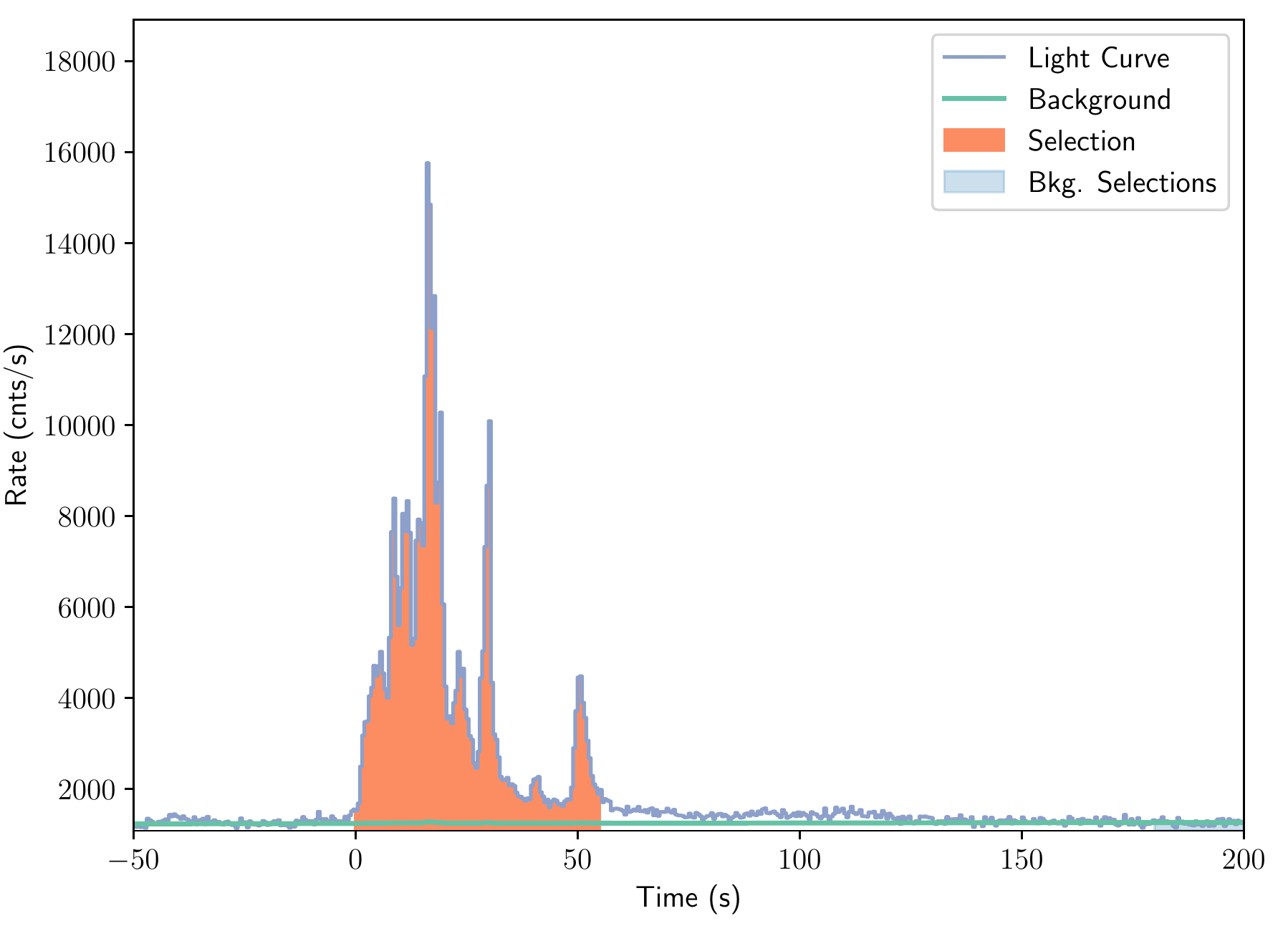}
B\includegraphics[width=0.471\hsize,clip]{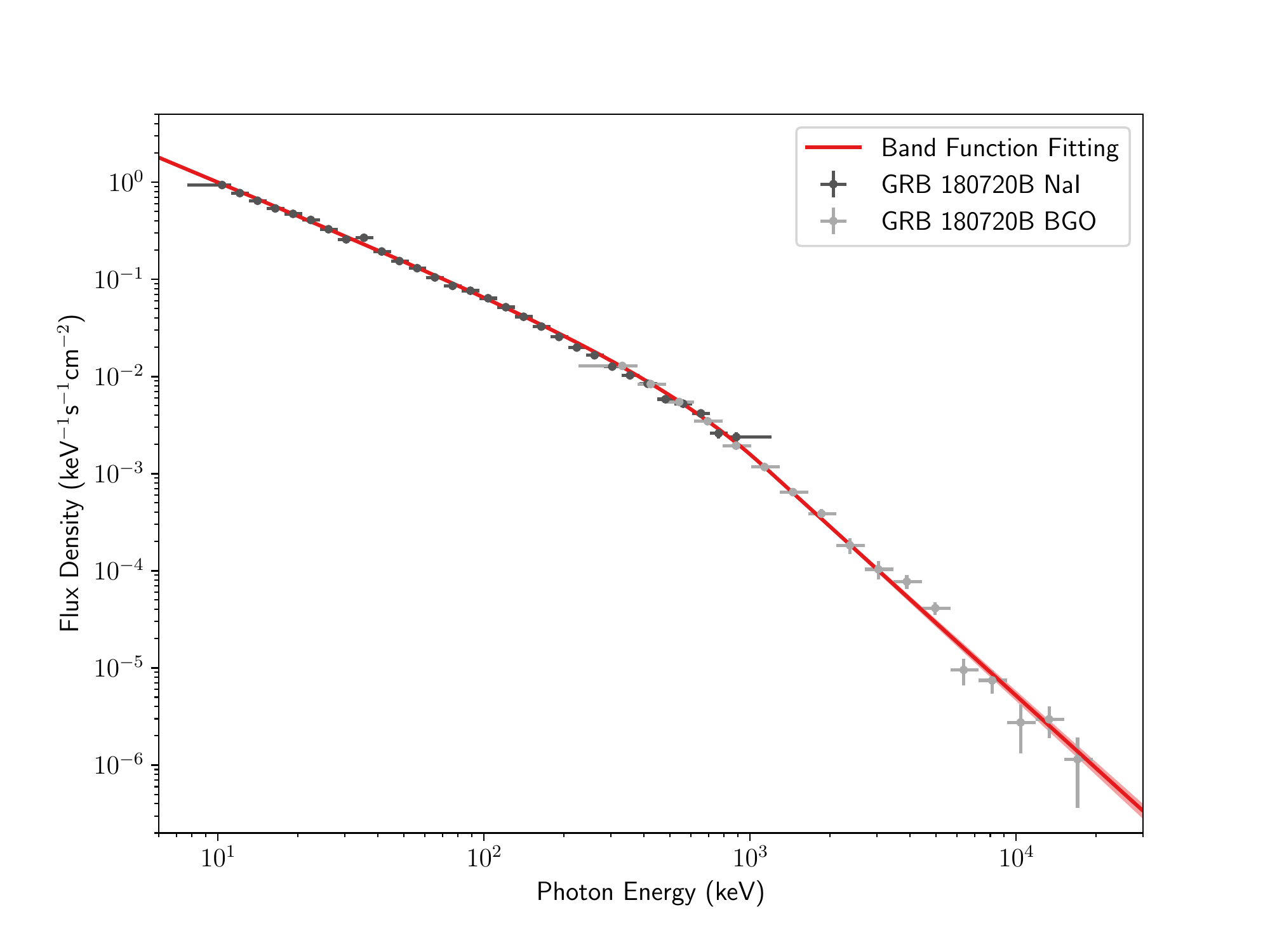}
\caption{\textbf{A:} Count rate of prompt emission: the count rate observation of NAI7 instrument on-boarded Fermi-GBM. The orange part from $0$ to $55$~s covers the $T_{90}$ duration. The entire emission extends to $\sim 120$~s. The green line is the fitted background. \textbf{B:} Spectrum of the prompt emission: it is very well fitted by a Band function during $0-55$~s, with the low-energy photon index $\alpha=-1.14$, the high-energy photon index $\beta=-2.48$, and the peak energy at $E_p = 712$~keV. The dark points are from NaI7 of Fermi-GBM, and the grey points are deduced from BGO of Fermi-GBM.}
\label{fig:180720B_prompt_rate}
\end{figure*}

The $T_{90}$ of GRB 180720B covers $0$--$55$~s in the observer's frame ($0$--$33$~s in the cosmological rest-frame) of Fermi-GBM and it contains multiple spiky structures; see Fig.~\ref{fig:180720B_prompt_rate}. The time integrated spectrum of $T_{90}$ was fitted by various functions, and compared by the deviance information criterion \citep{ando2010bayesian}, resulting that the best fit is given by the traditional Band function \citep{1993ApJ...413..281B}, with the peak energy $712$~keV, details in Fig.~\ref{fig:180720B_prompt_rate}.

The isotropic energy $E_{iso}$ from $1$~keV to $10$~MeV within the $T_{90}$ of this burst is $5.92 \times 10^{53}$~erg, where the cosmological expansion via the $k-$correction has been included \citep{2001AJ....121.2879B}.

\begin{figure*}
\centering
\includegraphics[angle=0, scale=0.7]{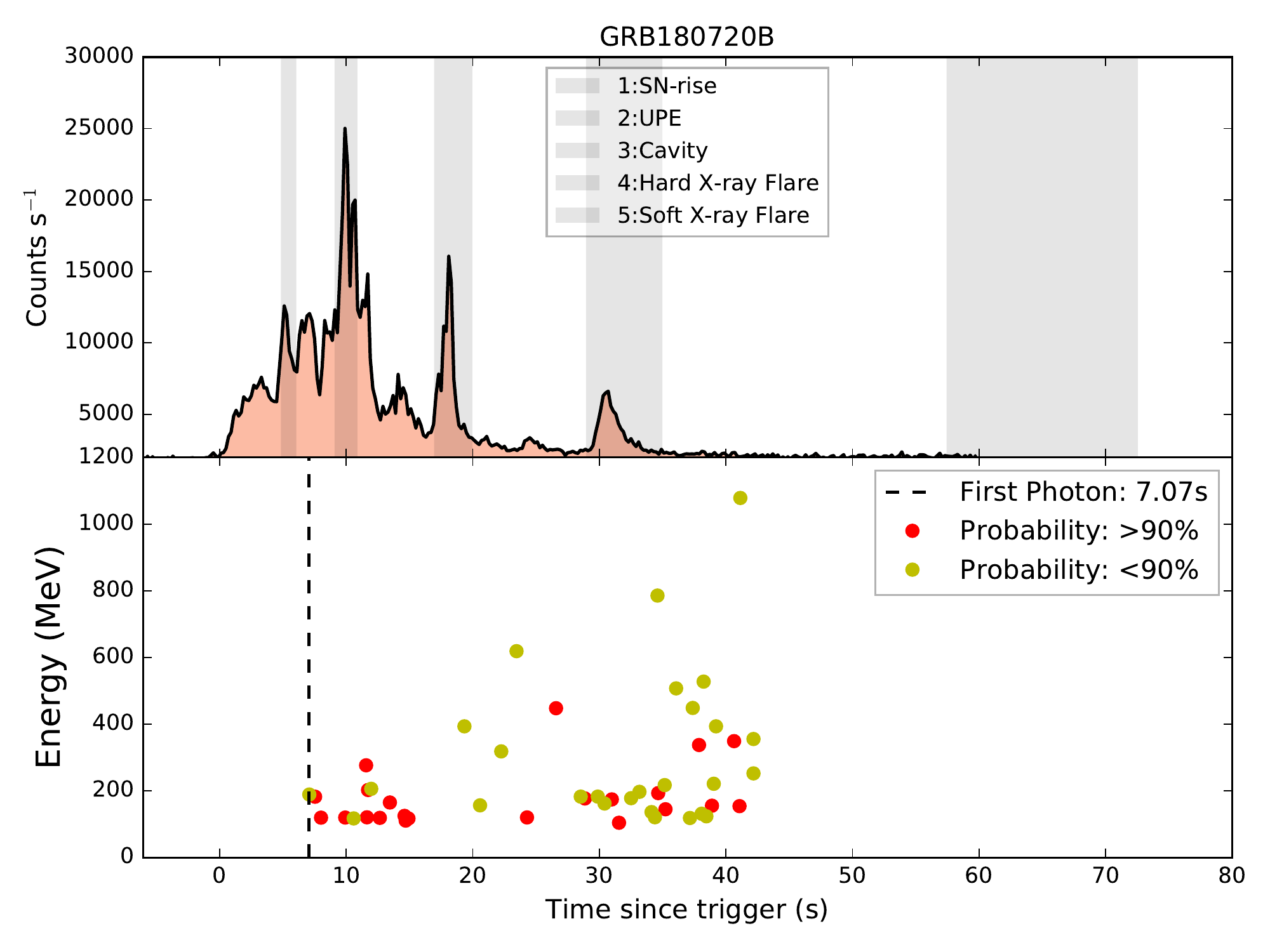}
\caption{Upper: Fermi-GBM light-curve. The five episodes of GRB 180720B are shown. Lower: The arrival time of GeV photons  as obtained from Fermi-LAT. Time is reported in the cosmological rest-frame of the source.}\label{fig:lc_180720B}
\end{figure*}

\begin{figure*}
A\includegraphics[angle=0,scale=0.42]{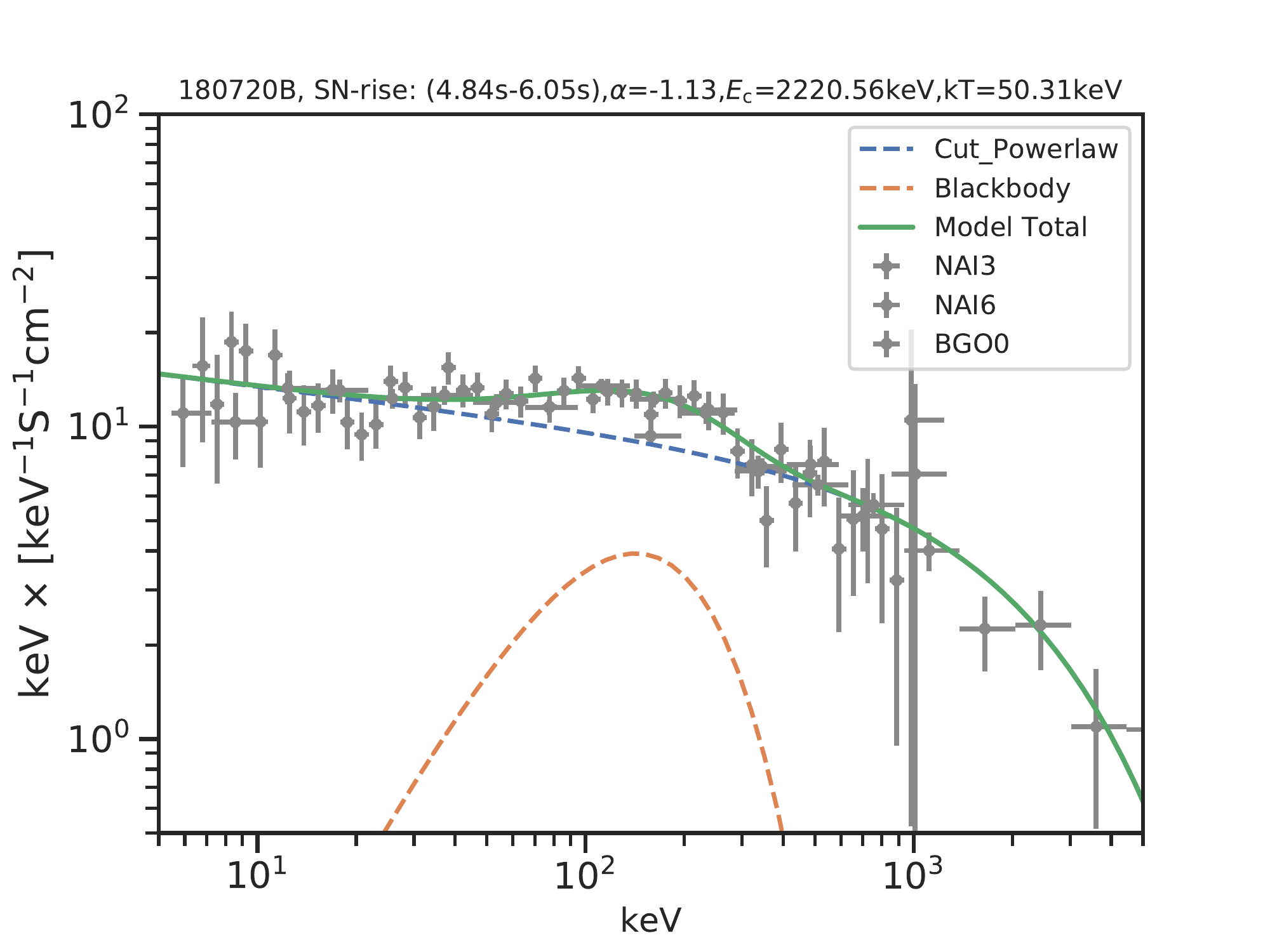}
B\includegraphics[angle=0,scale=0.42]{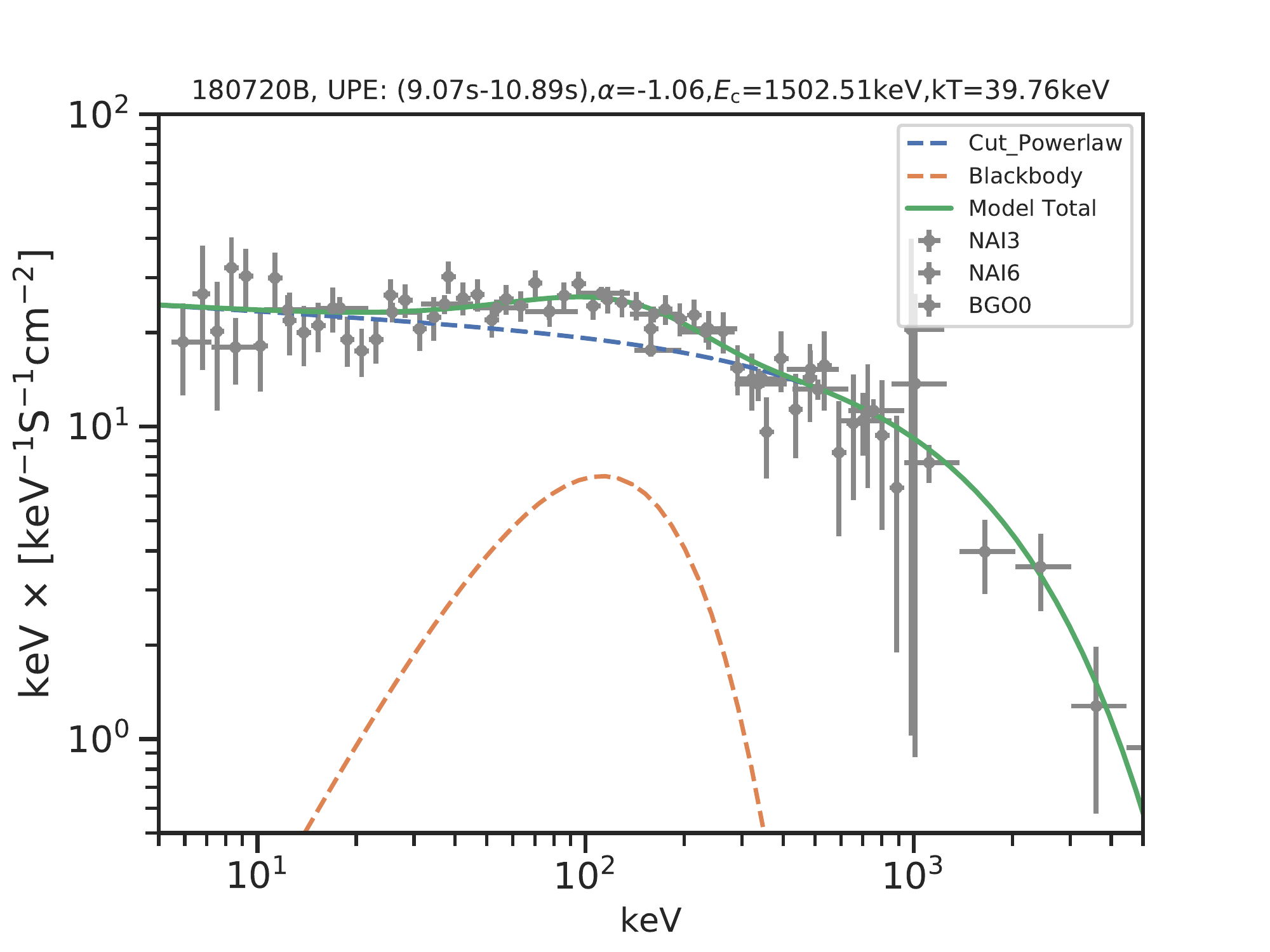}
C\includegraphics[angle=0,scale=0.42]{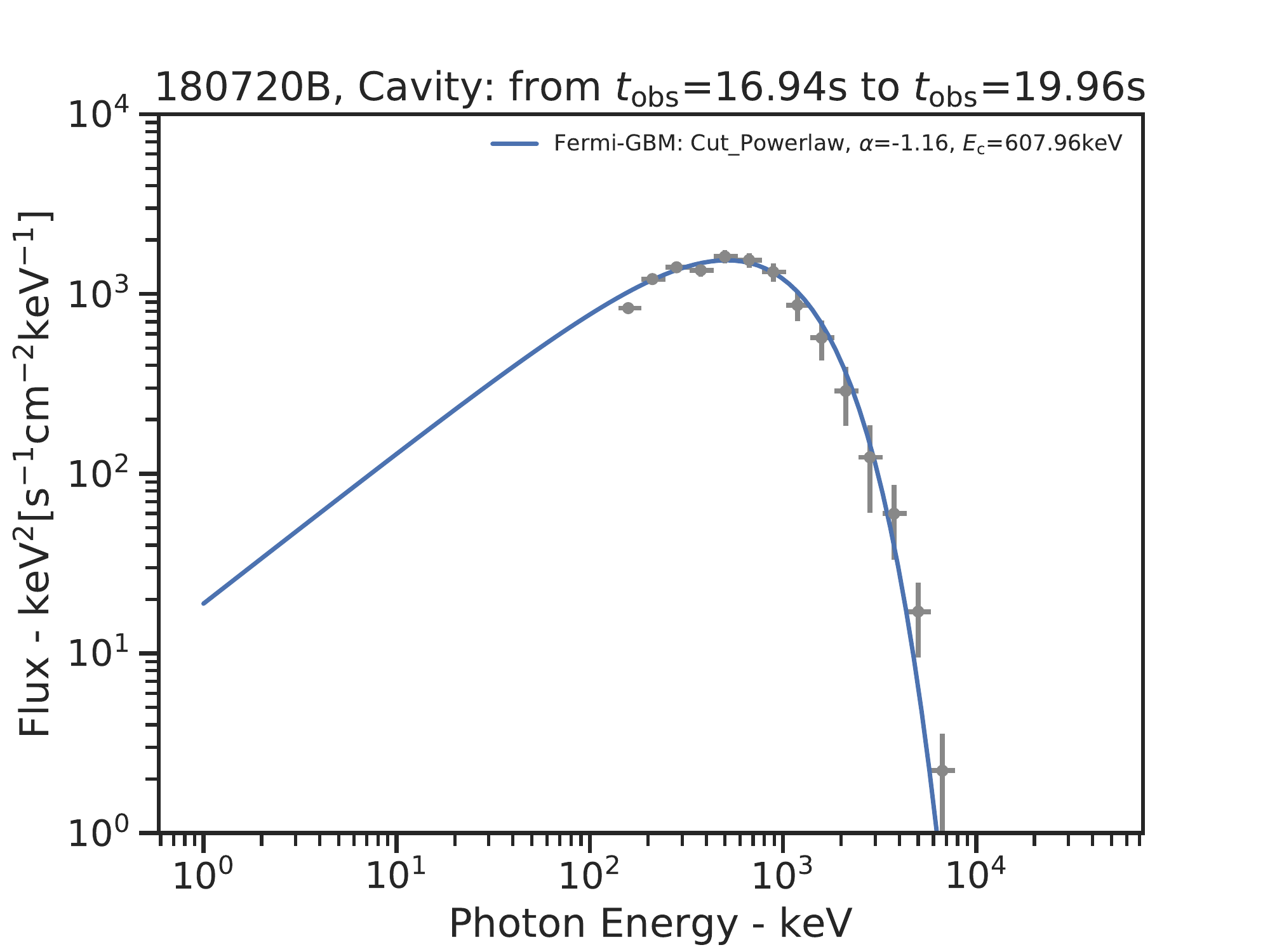}
D\includegraphics[angle=0,scale=0.42]{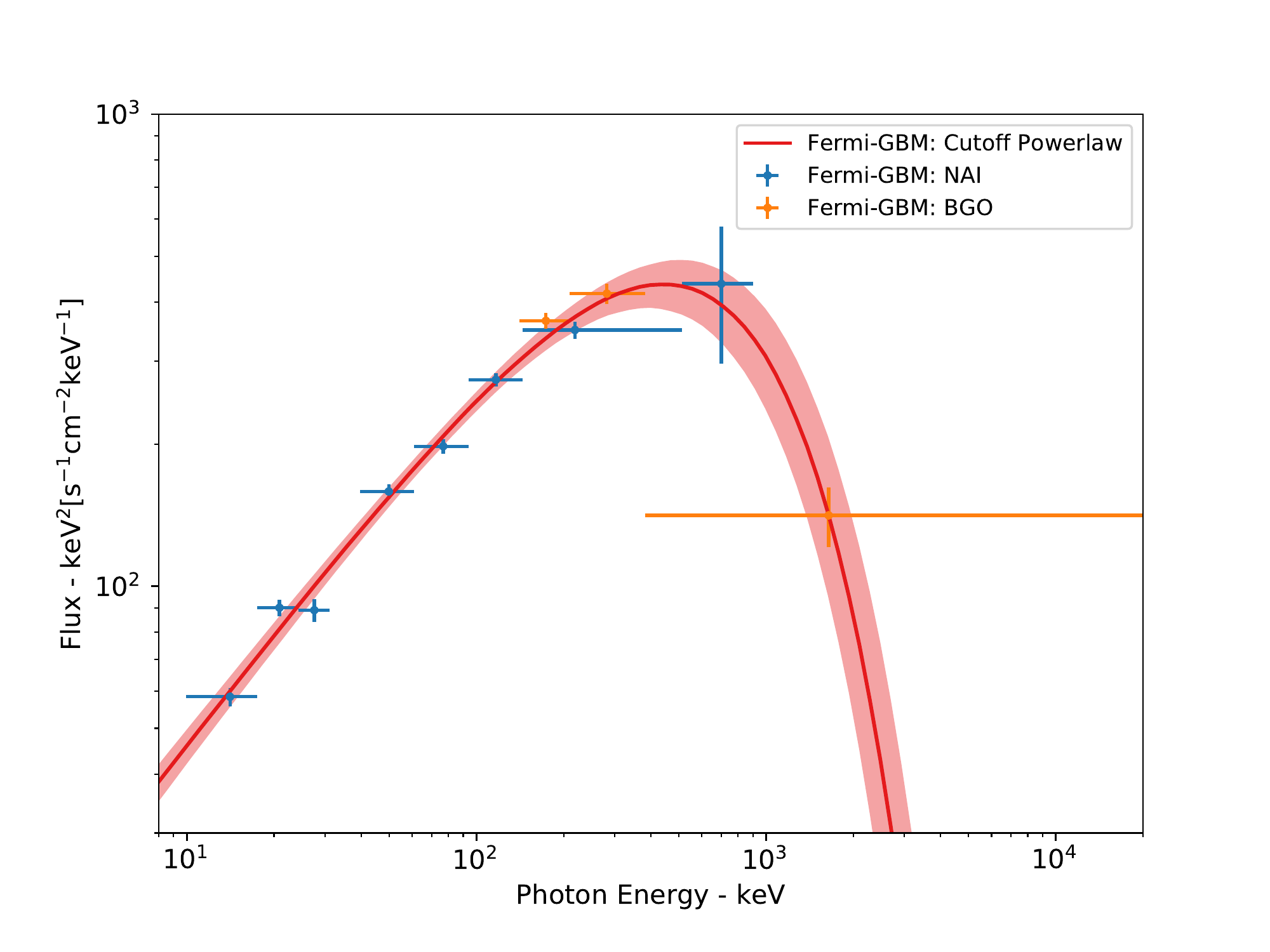}
E\includegraphics[angle=0,scale=0.31]{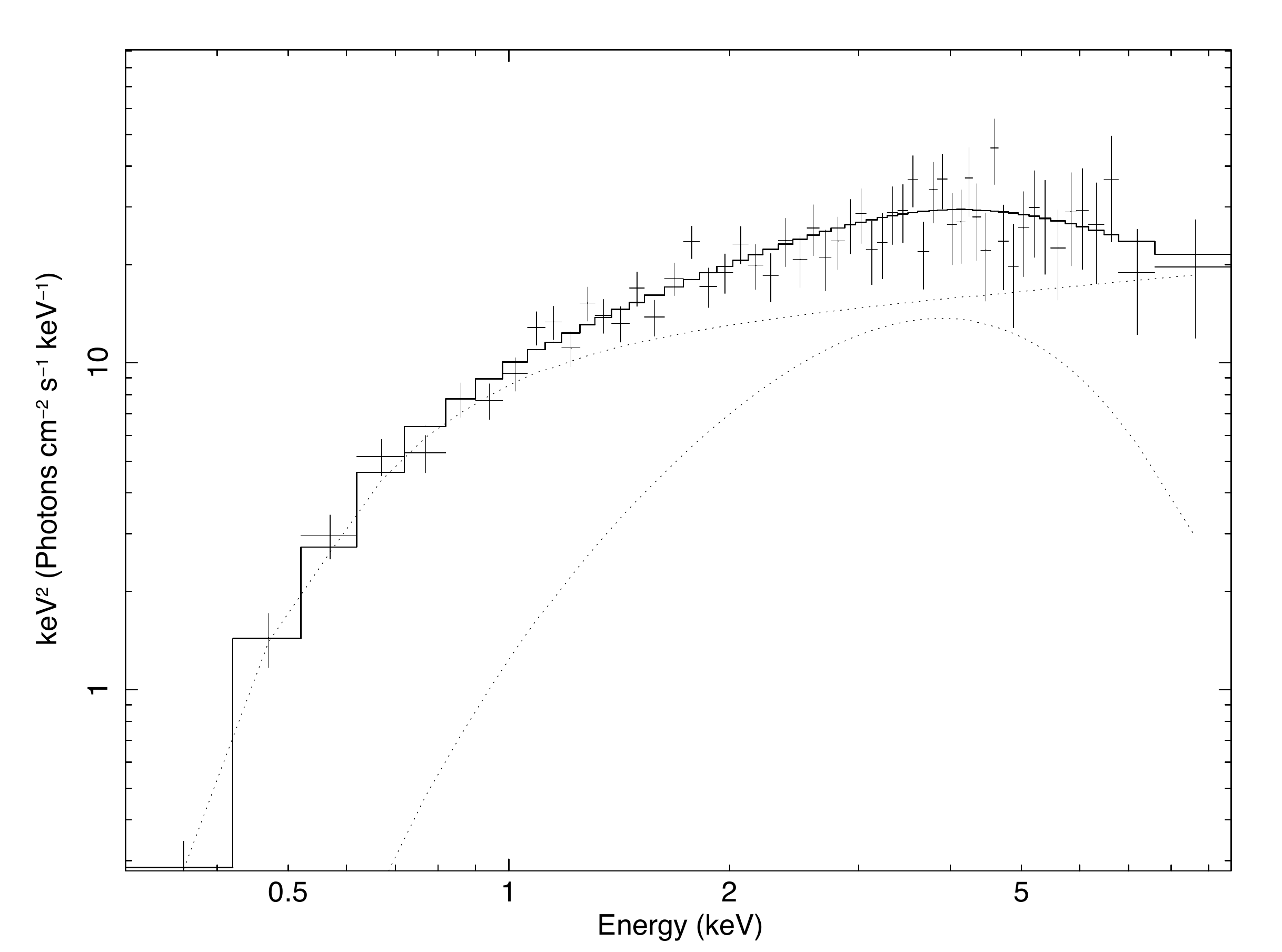}
\caption{The five episodes of GRB 180720B. \textbf{A:} \textit{SN-rise} spectrum fitted by a CPL+BB model from $t_{\rm rf}=4.84$ to $t_{\rm rf} = 6.05$s, spectral index $\alpha=-1.13$, cutoff energy $E_{\rm c}=2220.569$~keV, and BB temperature $k T = 50.31$~keV in the observer's frame. \textbf{B}: Spectrum analysis for $t_{\rm rf}=9.07$--$10.89$~s. The CPL+BB model is preferred with parameters: $\alpha= -1.06^{+0.01}_{-0.01}$, $E_{\rm c}=1502.5^{+88.6}_{-87.5}$~keV, and $kT= 39.8^{+1.6}_{-1.6}$~keV. The BB flux is $F_{\rm BB}=1.99^{0.43}_{-0.34}$ ($10^{-6}$~erg~cm$^{-2}$~s$^{-1}$) and the total one is $F_{\rm tot}= 45.55^{+3.11}_{-2.70}$ ($10^{-6}~$erg~cm$^{-2}$~s$^{-1}$). The BB to total flux ratio is $F_{\rm BB}/F_{\rm tot}=0.04$ and $E_{\rm iso}=1.6 \times 10^{53}$~erg (in this time interval). The best fit parameters are reported in Table~\ref{tab:180720B}. \textbf{C}: Spectrum of the \textit{cavity} in $t_{\rm rf} = 16.94$--$19.96$~s with parameters $\alpha=-1.16$, $E_{\rm c} = 607.96$~keV. There is no BB emission present indicating a ``featureless'' spectrum. \textbf{D:} HXF extending in $t_{\rm rf}= 28.95$--$34.98$~s, best fitted by a CPL model with $E_{\rm c}=(5.5_{-0.7}^{+0.8}) \times 10^2$~keV, $\alpha = -1.198 \pm 0.031$.  \textbf{E:} SXF extending in $t_{\rm rf}=57.43$--$72.55$~s with a PL+BB spectrum with $\alpha = -1.79 \pm 0.23$, and $k T=0.99 \pm 0.13$~keV.} \label{fig:3episodes}
\end{figure*}

We now perform a time-resolved spectral analysis of GRB 180720B in order to identify the different episodes of our model. From it, we have found five episodes with specific spectral features in the first $80$~s of the burst (see Fig.~\ref{fig:lc_180720B}), which we describe below.

\textbf{Episode 1}. This ``\textit{precursor}'' identified from $~t_{\rm rf}=4.84$~s to $~t_{\rm rf}=6.05$~s houses the \textit{SN-rise} of GRB 180720B. With an isotropic energy of $E^{\rm MeV}_{\rm SN}=(6.37\pm0.48) \times 10^{52}$~erg, it encompasses $15\%$ of the energy observed by Fermi-GBM. Its spectrum is best fitted by a CPL+BB model with photon index $\alpha=-1.13$, cutoff energy $E_{\rm c}=2220.569$~keV, and BB temperature $kT=50.31$~keV in the observer's frame; see Fig.~\ref{fig:3episodes} (A).

\begin{figure*}
\centering
\includegraphics[angle=0, scale=0.44]{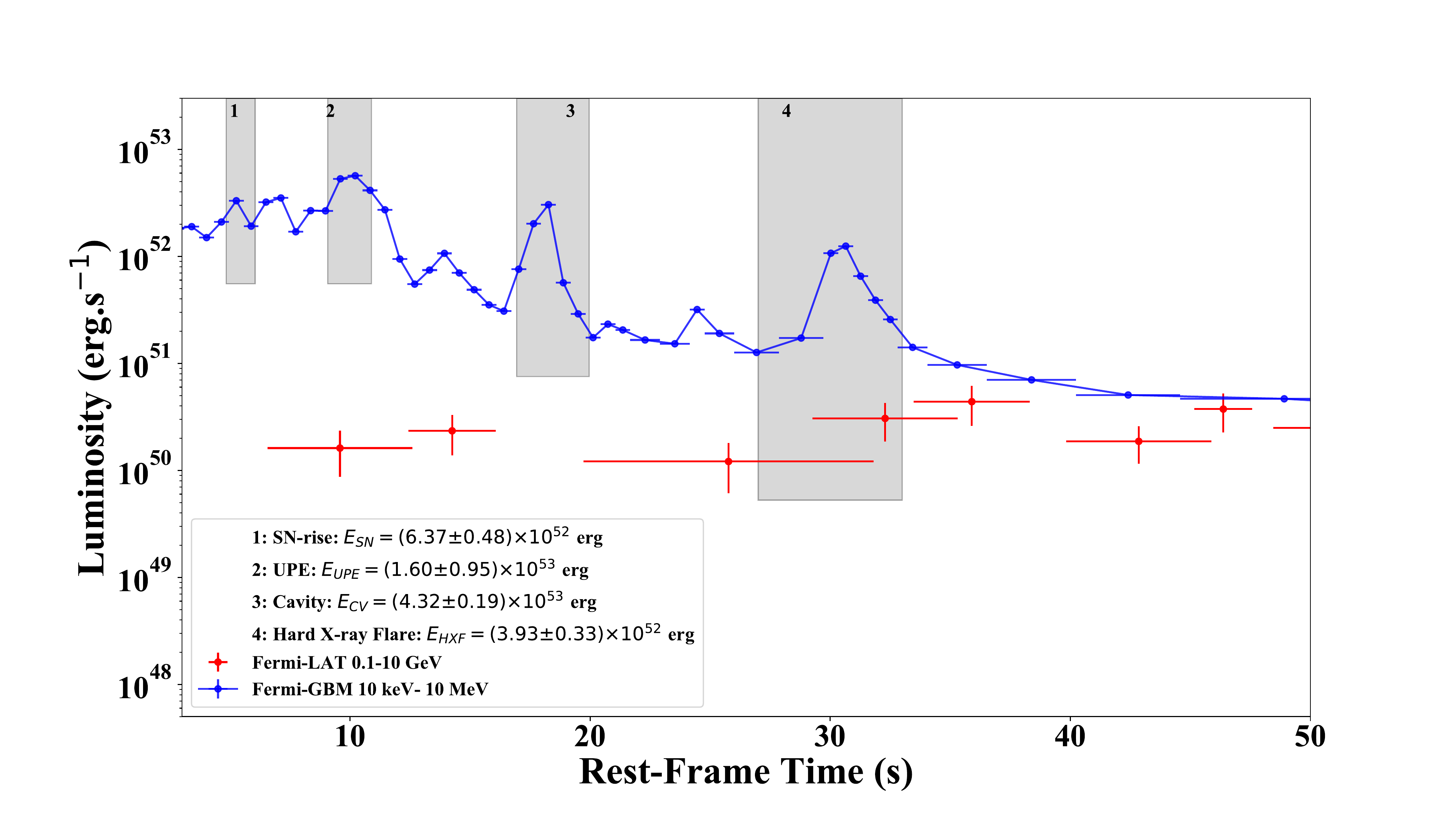}
\caption{Luminosity of the \textit{Fermi}-GBM in the $10$~keV--$10$~MeV energy band together with the luminosity of \textit{Fermi}-LAT in the $0.1$--$10$~GeV during the first 45 seconds, namely before the temporal decaying part of \textit{Fermi}-LAT luminosity, as expressed in the rest-frame of the source. In this time interval there are four episodes of GRB 180720B: Episode $1$ which contains the rise of the supernova (SN--rise) with energy of $E_{\rm SN}=(6.37 \pm 0.52)\times 10^{52}$~erg. Episode $2$, the ultra-relativistic emission phase (UPE)  with energy of $E^{\rm MeV}_{\rm UPE}=(3.6 \pm 1.1)\times 10^{53}$~erg. Episode 3, the cavity with energy of $E^{\rm MeV}_{\rm CV}=(1.08 \pm 0.07)\times 10^{53}$~erg and Episode 4, the hard X-ray flare (HXF) with energy of $E^{\rm MeV}_{\rm HXF}=(3.93\pm 0.33)\times 10^{53}$~erg. }\label{fig:1pdf}
\end{figure*}

\textbf{Episode 2}. This is the ``\textit{UPE phase}'' which lasts $1.82$~s, from $~t_{\rm rf}=9.07$~s to $~t_{\rm rf}=10.89$~s. With an isotropic energy of $E_{\rm UPE}^{\rm MeV}=(1.6 \pm 0.95) \times 10^{53}$~erg, it encompasses $36\%$ of the energy observed by Fermi-GBM; see Fig.~\ref{fig:3episodes} (B). This episode is best fitted by CPL+BB model with photon index $\alpha= -1.06^{+0.01}_{-0.01}$, cut-off energy $E_{\rm c}=1502.5^{+88.6}_{-87.5}$~keV, BB temperature $kT= 39.8^{+1.6}_{-1.6}$~keV. The BB flux is $F_{\rm BB}=1.99^{0.43}_{-0.34}$ ($10^{-6}$~erg~cm$^{-2}$~s$^{-1}$), and the total flux is $F_{\rm tot}=$ 45.55 $^{+3.11}_{-2.70}$ ($10^{-6}~$erg~cm$^{-2}$~s$^{-1}$). The BB to total flux ratio is, therefore, $F_{\rm BB}/F_{\rm tot}=0.04$. The time-resolved spectral analysis of this Episode is presented in Appendix~\ref{sec:time-resolved}.

The physical interpretation of UPE phase lasting only $2$~s here, has been addressed in the case of GRB 190114C (Moradi et al, submitted). There the UPE phase characterizes a well--defined Episode lasting also only $2$~s. There we show how the physical origin of the MeV radiation observed in the UPE phase originates in an overcritical regime of the \textit{inner engine}, creating an $e^{\pm}~\gamma$ plasma in presence of a baryon load. What has been tested in the BdHN model is the existence of some Episodes which can be independently described and they do not affect the sequence of other Episodes. 

\textbf{Episode 3}; hosts the \textit{cavity} from $t_{\rm rf}=16.94$~s to $~t_{\rm rf}=19.96$~s, with an equivalent isotropic energy of $E_{\rm CV}^{\rm MeV}=(4.32 \pm 0.19) \times 10^{52}$~erg. It encompasses 10$\%$ of the energy observed by Fermi-GBM. Its spectrum is best fitted by a CPL model with photon index $\alpha= -1.16$ and the cutoff energy $E_{\rm c}=607.96$~keV. No BB component is present in the spectrum, indicating a featureless spectrum; see Fig.~\ref{fig:3episodes} (C). This is in total analogy with the observations of the cavity of GRB 130427A \citep{2019ApJ...886...82R} and GRB 190114C \citep{2019ApJ...883..191R,2019arXiv190404162R}.

Following these three Episodes, there are two additional episodes in the early afterglow phase:

\textbf{Episode 4}. This is the hard X-ray flare (HXF) which starts from $t_{\rm rf}= 28.95$~s and ends at $t_{\rm rf}= 34.98$~s, with an isotropic energy of $E_{\rm HXF}^{\rm MeV}=(3.93 \pm 0.33) \times 10^{52}$~erg. It contains $7\%$ of the energy observed by Fermi-GBM. The luminosity lightcurve is shown in Figs.~\ref{fig:GBM_BAT_Linear} and \ref{fig:1pdf}. The spectrum is best fitted by a CPL model with photon index of $\alpha=-1.198 \pm 0.031$ and a cut-off energy $E_{\rm c} = 5.5_{-0.7}^{+0.8}) \times 10^2$~keV.

\textbf{Episode 5}. This is the soft X-ray flare (SXF) with isotropic energy of $2.89\times 10^{51}$~erg in $0.3$--$10$~keV band as observed by Swift-XRT. It starts from $t_{\rm rf}=57.43$~s and ends at $t_{\rm rf}= 72.55$~s. The luminosity lightcurve is shown in Figs.~\ref{fig:GBM_BAT_Linear} and \ref{fig:3pdf}. The spectrum is best fitted by a power-law plus a blackbody component (PL+BB) with a power-law index of $-1.79 \pm 0.23$, and a BB temperature of $0.99 \pm 0.13$~keV.

In addition to above five episodes in the first $80$~s of GRB 180720B, the $0.1$--$10$~GeV emission observed by Fermi-LAT is also present. The first photon with probability more than $90\%$ belonging to GRB 180720B is a $\sim 200$~MeV event observed at $t_{\rm rf}= 7.07$~s after the GBM trigger; see Fig.~\ref{fig:1pdf}. The highest-energy photon is a $5$~GeV event observed at $t_{\rm rf}=83$~s after the GBM trigger \citep{2018GCN.22980....1B}. 

\begin{figure*}
\centering
\includegraphics[angle=0, scale=0.44]{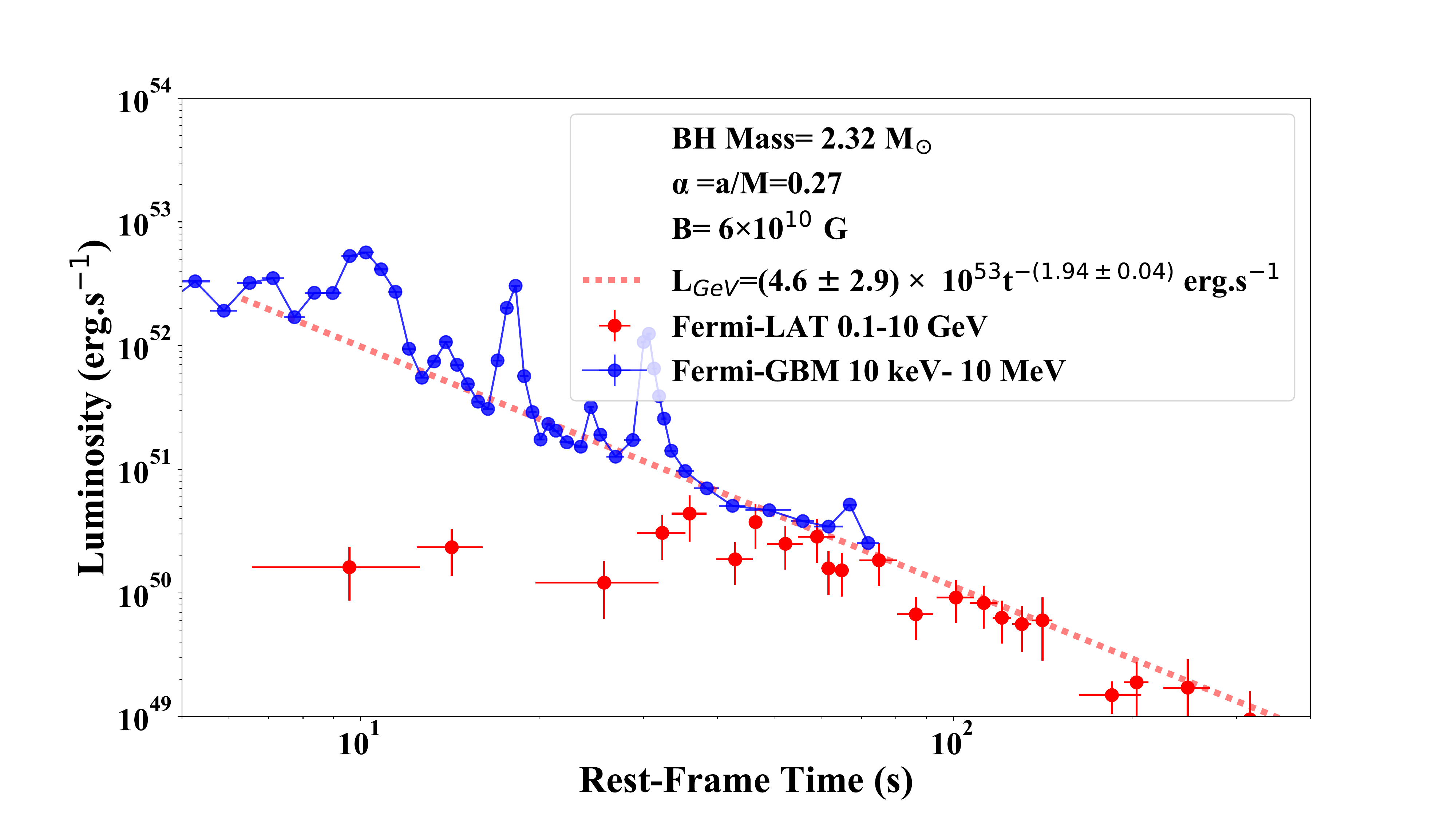}
\caption{\textbf{a:} Luminosity of the \textit{Fermi}-GBM in the $10$~keV--$10$~MeV energy band together with the luminosity of \textit{Fermi}-LAT in the $0.1$--$10$~GeV as expressed in the rest-frame of the source. The red dashed line shows the extrapolation of the GeV best fit after $45$~s back to the times before $45$~s. The anti-correlation between the luminosity obtained from \textit{Fermi}-GBM data and \textit{Fermi}-LAT data is obvious. The mass and spin of the Kerr BH as inferred from the GeV radiation are $M=2.32 M_{\odot}$ and $\alpha=0.27$, respectively.}\label{fig:2pdf}
\end{figure*}

Contrary to the MeV luminosity, during the Episodes $2$--$4$, namely in the time interval $t_{\rm rf}=7.07$--$40$~s, the GeV luminosity shows an increasing behavior with time; see Fig.~\ref{fig:1pdf}. There are two possible mechanisms that might explain the anti-correlation between the luminosity of \textit{Fermi}-GBM and \textit{Fermi}-LAT in this time interval. The GeV photons can be absorbed by the MeV photons via the the Breit-Wheeler two-photon pair creation process \citep{1934PhRv...46.1087B}, $\gamma+\gamma\rightarrow e^{\pm}$ \citep[see also][]{2010PhR...487....1R, 2016Ap&SS.361...82R}, or the magnetic field can be screened by the $e^{\pm}$ pairs (S. Campion, et al., submitted), thereby decreasing the opacity to the MPP process \citep{2019ApJ...886...82R}. In either scenario, the GeV luminosity increases up to the point of transparency where it starts to show its intrinsic behavior of a decreasing power-law, see Figs.~\ref{fig:1pdf} and \ref{fig:2pdf}. 

Considering the existence of these $5$ Episodes, GRB 18072B is similar to GRB 190114C \citep{2019arXiv190404162R}, with the difference that GRB 190114C is observed with a viewing angle normal to the plane of the binary progenitor, thereby explaining the lack of observed HXF and SXF. In Table~\ref{tab:Comparison}, we present the similarities of GRB 180720B and GRB 190114C.

\begin{table*}[!ht]
\centering
\caption{The presence of different Episodes in GRB 180720B and GRB 190114C. GRB 190114C has the viewing angle normal to the binary plane, therefore there is no prominent observed HXF and SXF.}
\label{tab:Comparison}
\small
\begin{tabular}{@{}cccccccc@{}}

\hline
&Characteristic~&~~GRB 180720B~~&~~GRB 190114C\\
&&~~ ~~&~~\citep{2019arXiv191012615L}~\\
&&~~ ~~&~~\citep{2019arXiv190107505W}~~\\
\hline\hline 
&~~Duration (s)~~&1.21&1.9&\\
~~Episode 1 (SN--rise) ~~&&&\\
 &Energy (erg)&~$(6.37\pm 0.48) \times 10^{52}$~~~&~$(1.00\pm 0.12) \times 10^{53}$~&\\
 
  \hline
&~~Duration (s)~~&1.82&2.09\\
~~Episode 2 (UPE) ~~&&&\\
 &Energy(erg)&~$(1.6\pm 0.95) \times 10^{53}$~~~&~$(1.47\pm 0.20) \times 10^{53}$~\\

  \hline
  &~~Duration (s)~~&3.02&13.1\\
~~Episode 3 (Cavity) ~~&&&\\
 &Energy(erg)&$(4.32\pm 0.19) \times 10^{52}$~~~&~$(2.49\pm 0.12) \times 10^{52}$~~~\\
  \hline
  &~~Duration (s)~~&6.03&--\\
~~Episode 4 (HXF) ~~& &&No HXF&\\
 &Energy(erg)&$(3.93\pm 0.33) \times 10^{52}$~~~&\\
  \hline
  &~~Duration (s)~~&15.12&--\\
~~Episode 5 (SXF) ~~&&&No SXF\\
 &Energy(erg)&$(2.89\pm 042) \times 10^{52}$~~~&\\
\hline 
\end{tabular}
\end{table*}

\section{Data analysis after the $T_{\rm 90}$ of GRB 180720B}\label{sec:aftert90}

\subsection{The X-ray Afterglow\label{sec:x_ray}} 

At 14:23:11.0 UT, namely about $86$~s ($52$~s in the cosmological rest-frame time) after the BAT trigger, the Swift-XRT started to observe the source. XRT detected a flaring and fading X-ray source at the beginning of the observation and decaying power-low flux at times later than $\sim 1000$~s \citep{2018GCN.22975....1S}.

\begin{figure*}
\centering
\includegraphics[angle=0, scale=0.9]{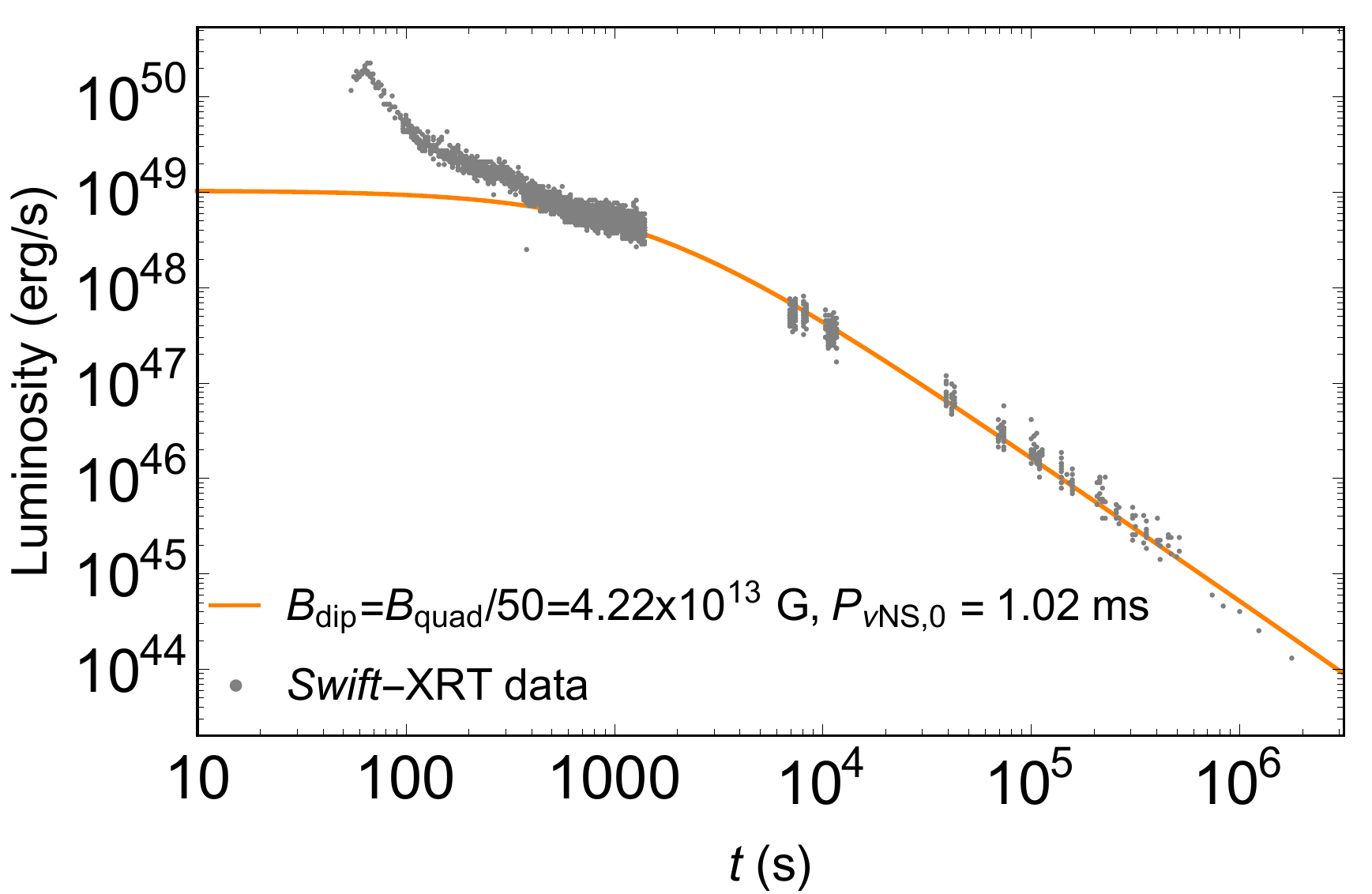}
\caption{The black points correspond to the bolometric (about $\sim 5$ times brighter than the soft X-ray observed by Swift-XRT) light-curves of GRB 180720B. The solid line is the fitting of the energy injection from the rotational energy of the $\nu$NS. Till $\sim 5 \times 10^2$~s the mildly-relativistic SN kinetic energy \citep{2018ApJ...852...53R, 2018ApJ...869..151R} and the BH are responsible for the first five episodes of GRB 180720B. After $ \sim 5 \times 10^2$~s, the activity of the $\nu$NS is dominant in the radiation process \citep{2018ApJ...869..101R}. The fitting parameters are shown in the legend and of Table \ref{tab:ns_parameters}, the strength of the quadruple field is given in a range, its upper value is three times the lower value due to the oscillation angle $\chi_2$, which is a free parameter (see \citealp{2019ApJ...874...39W}, for details).}\label{fig:lumthNS}
\end{figure*}

\begin{figure*}
\centering
\includegraphics[angle=0, scale=0.44]{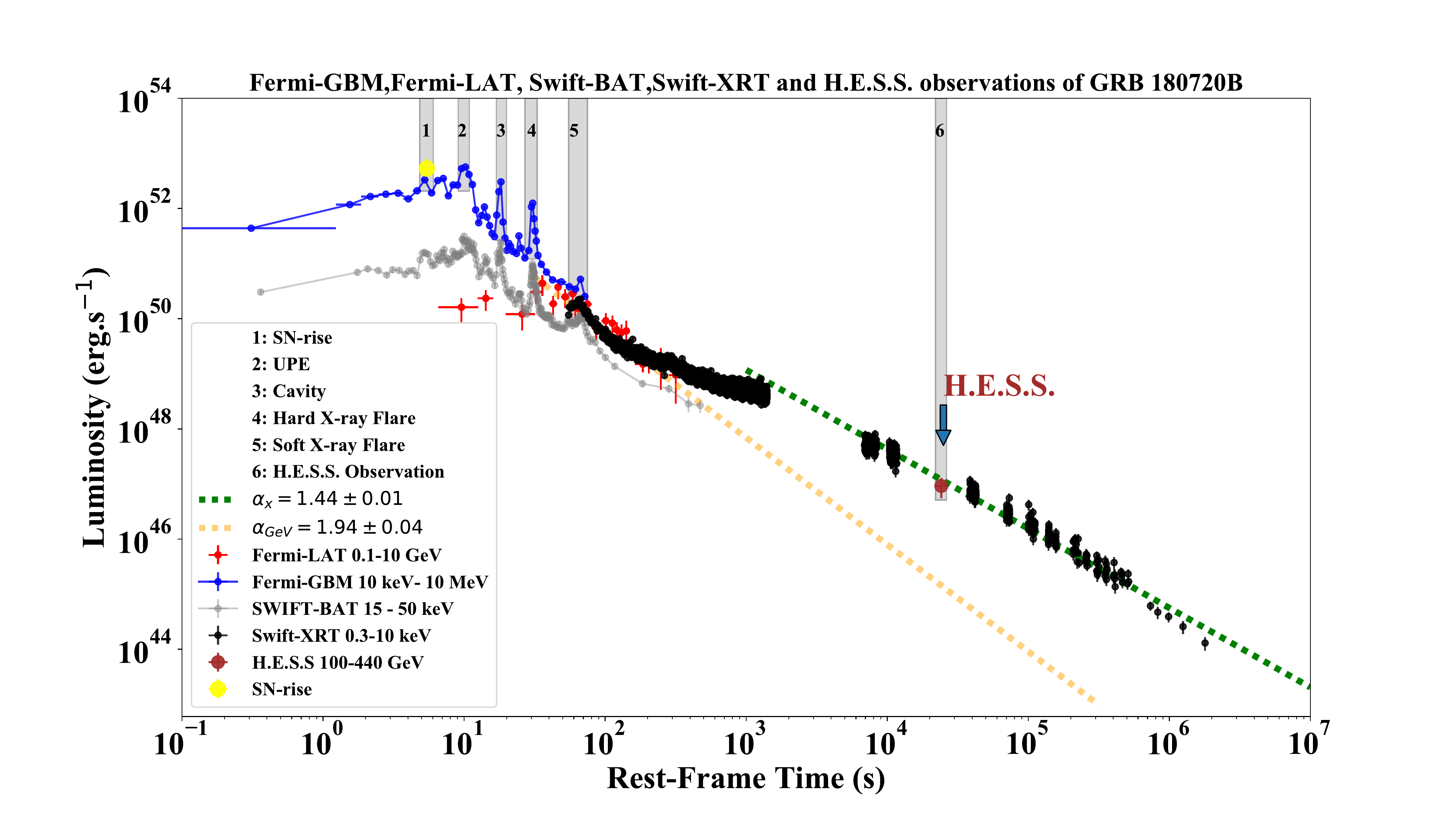}
\caption{Luminosity light-curves obtained from Fermi-GBM, in $10$~keV--$10$~MeV, Fermi-LAT in $0.1$~GeV--$10$~GeV,Swift-BAT in $15$~keV--$50$~keV, Swift-XRT in $3$~keV--$10$~keV, and H.E.S.S. in $100$~GeV--$440$~GeV. The late X-ray afterglow luminosity of BdHN I GRB 180720B observed by \textit{Swift}-XRT is best fit by a temporal decaying power law of $L_{\rm X}= (2.5\pm 0.4)\times 10^{53} \rm ~t^{1.44\pm 0.01} erg.s^{-1}$.  The light-curve of Fermi--LAT in is fitted by temporal decaying power law of $L_{\rm GeV}= (4.6\pm 2.9)\times 10^{53} \rm ~t^{-1.94\pm 0.04} erg.s^{-1}$, clearly different from the fit of X-ray afterglow. The different in the temporal decay of luminosity in GeV and X-ray band indicates their different radiation source.}\label{fig:3pdf}
\end{figure*}

Figures~\ref{fig:lumthNS} and \ref{fig:3pdf} show the luminosity of the ``flare-plateau-afterglow'' (FPA) phase observed by the Swift-XRT in the $0.3$--$10$~keV band. It starts by a flare at rest-frame time $t_{\rm rf}\sim 52$~s, the light-curve continues with a semi-plateau, shallow decay till $10^3$~s, then it follows the normal decay phase, shown in Fig.~\ref{fig:3pdf}. The soft X-ray energy observed from $50$~s to $10^6$~s is $4.03 \times 10^{52}$~erg.

The presence of flaring activities in the early afterglow phase of this GRB indicates the viewing angle of observation lying in the equatorial plane of the binary progenitor. An example of such a viewing angle has been identified in GRB 151027A by \citet{2018ApJ...869..151R}. Further examples of such cases have been shown in \citet{2018ApJ...852...53R}. However, since GeV radiation has been detected by Fermi-LAT for this GRB (see section~\ref{sec:G_ray} for details), the viewing angle has to be somewhere between the rotation axis and the equatorial plane of the progenitor system \citep[see the case of BdHN I 130427A in][as an example of the viewing angle lying on the rotation axis]{2019ApJ...886...82R}.

The $0.3$--$10$~keV luminosity light-curve of GRB 180720B at t~$\gtrsim 10^4$~s follows a decaying power-law:
\begin{equation}\label{LX}
    L_{\rm X} =A_{\rm X}~\left(\frac{t}{{1\rm s}}\right)^{-\alpha_{\rm X}},
\end{equation}
with amplitude $(2.5\pm 0.4)\times 10^{53} \rm erg.s^{-1}$ and an index of $\alpha_X=1.44\pm 0.01$; see Fig.~\ref{fig:GeV} (b) and  Fig.~\ref{fig:3pdf}.

In \citet{2018ApJ...869..101R} and in this article in section~\ref{sec:nuns}, it is described how the X-ray afterglow originates from the synchrotron radiation of relativistic electrons in the mildly-relativistic expanding magnetized HN ejecta due to the presence of the \textbf{$\nu$NS}. 

\subsection{The GeV Radiation of GRB 180720B} 
\label{sec:G_ray}

We have analyzed the data of Fermi-LAT using the \textit{gtburst} interface which is a free interface as part of the official \textit{Fermitools} software. Figures~\ref{fig:GeV} (a) and \ref{fig:2pdf} show our best fits for the luminosity of the GeV emission in its temporal decaying part, observed by Fermi-LAT in $0.1$--$10$~GeV band. Data points are fitted by the LMFIT package of Python which uses Non-Linear Least-Squares Minimization. 

The $0.1$--$10$~GeV emission of GRB 180720B after the $\sim $40~s has a luminosity that satisfy
\begin{equation}\label{L3}
    L_{\rm GeV} =A_{\rm GeV}~\left(\frac{t}{{1\rm s}}\right)^{-\alpha_{\rm GeV}},
\end{equation}
with amplitude $(4.6\pm 2.9)\times 10^{53}$~erg~s$^{-1}$ and a slope of index of $\alpha_X=1.94\pm 0.13$; see Fig.~\ref{fig:GeV} (a) Fig.~\ref{fig:2pdf}.

The GeV radiation data observed by Fermi-LAT allows the determination of the BH mass \citep{2019ApJ...886...82R} and the X-ray afterglow data observed by the \textit{Swift}-XRT allows to determine the spin and magnetic field structure of the $\nu$NS \citep{2018ApJ...869..101R}. In the next sections we first present the properties of the BH as a component of the \textit{inner engine} of GRB 180720B and then the ones of $\nu$NS. 

\subsection{H.E.S.S. observations}

At $T_0 + 10.1$~h ($t_{\rm rf}=21956.5$~s) the observations with H.E.S.S. started and lasted for about $\sim 1.2$~h ($t_{\rm rf}=26304.3~s$), revealing a new $\gamma$-ray source with an excess of $119$ $\gamma$-ray events with a statistical significance of $5.3\sigma$ \citep{2019Natur.575..464A}. The isotropic luminosity in this time interval is $L^{\rm iso}_{\rm HESS}= (5.5\pm 4.2)\times 10^{46}$~erg~s$^{-1}$, with the corresponding isotropic energy of $E^{\rm iso}_{\rm HESS}= (2.4\pm 1.8) \times 10^{50}$~erg. We call this phase as a new Episode $6$ shown in Fig.~\ref{fig:GeV} (c) and  Fig.~\ref{fig:3pdf}.

The GRB position was re-observed  by H.E.S.S. $18$~d after the first observation for $6.75$~h. Results were consistent with background events. This shows that observation done by H.E.S.S. is not associated with an unknown steady Gamma-ray emitter. 

\section{Physical origins of the GeV, the X-ray afterglow, and the H.E.E.S. radiations}\label{sec:BH-NS-Hess}

\subsection{The \textit{inner engine} of GRB 180720B and the mass and spin of the BH}\label{sec:inner}

It has been shown that the nature of the GeV emission can be explained by an electro-gravitomagnetic process that extracts the rotational energy of the Kerr BH \citep{2019ApJ...886...82R}. This mechanism works when the Kerr BH is placed in an external uniform magnetic field, parallel to the rotation axis as described by the Papapetrou-Wald solution \citep{1966AIHPA...4...83P, 1974PhRvD..10.1680W}, and immersed in a very-low-density fully ionized plasma. Consequently, the formation of the BH coincides with the onset of the GeV emission revealing itself by the observation the first GeV photon by Fermi-LAT. This framework has led to determination of the mass and spin of the BH and the strength of the surrounding magnetic field. 

The first GeV photon of the GRB 180720B with probability more than $90\%$ belonging to this GRB is a $\sim 200$~MeV observed at $t_{\rm rf}= 7.07$~s after the GBM trigger; see Fig.~\ref{fig:lc_180720B}. The onset of GeV radiation is $2$~s prior to the onset of the UPE phase, indicating that the surrounding magnetic field at the moment of the formation of BH is not strong enough to produce the critical electric field around the BH. The amplification of the magnetic field has been shown in the numerical solution of the massive NS, collapsing to the BH endowed with a nearly uniform external magnetic field and its accretion disk \citep[see][for details]{2006PhRvL..96c1101D,2006PhRvL..96c1102S,2006PhRvD..73j4015D,2007CQGra..24S.207S,2008PhRvD..77d4001S}. 

As an indicative example, the three-dimensional numerical simulation done by \citet{2011ApJ...732L...6R} show the amplification of the magnetic field from $\sim 10^{12}$~G to $\sim 10^{15}$~G in such a scenario. Following these numerical simulations, \citet{2020ApJ...893..148R} have proposed the post-merger analogous in BdHNe I relating the process to the newborn BH. Therefore, it is expected that after the formation of the BH, due to the accretion process, the magnetic field becomes stronger and after some seconds, for GRB 180720B two seconds, it becomes overcritical, and triggers the UPE phase. 

We now apply the self-consistent method utilizing Papapetrou-Wald solution, already well tested in the case of GRB 130427A \citep{2019ApJ...886...82R} and GRB 190114C \citep{2019arXiv191107552M} for determining the three parameters of the \textit{inner engine}, namely the mass and spin of the BH, and the strength of the surrounding magnetic field, $B_0$. The values are obtained satisfying three conditions:
\begin{itemize}
\item  
The magnetic field strength is such that the GeV photons are not subjected to the magnetic $e^+e^-$ pair production (MPP) process, therefore the transparency condition is fulfilled.
\item 
The energy of the GeV photons, produced by synchrotron radiation, reduces by an equal amount the mass-energy of the BH, i.e. the GeV emission is powered by the Kerr BH rotational energy.
\item  
The timescale of the synchrotron radiation equals the observed GeV radiation timescale.
\end{itemize}

With these conditions, we obtain the lower limit of the mass, $M=2.32 M_\odot$, and spin parameter of the BH, $\alpha = 0.27$. The corresponding constant irreducible mass of the BH is $M_{\rm irr}=2.29~M_\odot$. The strength of the surrounding magnetic field is $\approx 6\times 10^{10}$~G; see Fig.~\ref{fig:1pdf}.

\subsection{The spin and magnetic field of the  $\nu$NS}\label{sec:nuns}

In \citet{2018ApJ...869..101R} and \citet{2019ApJ...874...39W}, the bolometric luminosity from the $\nu$NS rotational energy loss by magnetic braking has been modeled for the emission at late times $t\gtrsim 10^3$~s of the ``Nousek-Zhang'' \citep{Nousek2006,Zhang2006} (flare-plateau-afterglow, FPA phase). From this model, we have inferred the rotation period of the $\nu$NS as well as its magnetic field structure \citep{2020ApJ...893..148R}.

Following \citet{2018ApJ...869..101R, 2019ApJ...874...39W, 2020ApJ...893..148R}, we here assume that the $\nu$NS rotational energy, $E_{\rm rot}$, is responsible for the observed electromagnetic emission of the X-ray afterglow, $E_{\rm X}$, i.e.:
\begin{equation}\label{eq:ErotEx}
	E_{\rm rot} = \frac{1}{2} I \Omega^2=E_{\rm X},
\end{equation}
where $\Omega = 2\pi/P_{\nu\rm NS}$ is the $\nu$NS angular velocity and $I$ its the moment of inertia. For instance, a $\nu$NS with $I=10^{45}$~g~cm$^2$ and initial rotation period $P_{\nu\rm NS}=~1$~ms, has a rotational energy $E_{\rm rot} \approx 2 \times 10^{52}$~erg. We assume that the rotational energy loss is driven by magnetic braking provided by the dipole and quadruple components, i.e.:
\begin{multline}
L_{\rm X}=L_{\nu \text{NS}}(t) = -\dot{E}_{\rm rot}= -I \Omega  \dot{\Omega } \\
    = - \frac{2}{3 c^3} \Omega^4 B_{\rm dip}^2 R_{\nu\rm NS}^6 \sin^2\chi_1  \left(1+\eta^2 \frac{16}{45} \frac{R_{\nu\rm NS}^2 \Omega^2}{c^2}\right),
\end{multline}\label{eq:pulsar_luminosity}
where
\begin{equation}
    \eta^2 = (\cos^2\chi_2+10\sin^2\chi_2) \frac{B_{\rm quad}^2}{B_{\rm dip}^2},
\label{eq:eta}
\end{equation}
measures the relative strength of the quadruple and dipole components, and an overdot stands for time derivative. The parameters $\chi_1$ and $\chi_2$ are the inclination angles of the magnetic moments and $B_{\rm dip}$ and $B_{\rm quad}$ are, respectively, the strengths of the dipole and quadruple magnetic field components. 
\begin{figure}
\centering
\includegraphics[width=\hsize,clip]{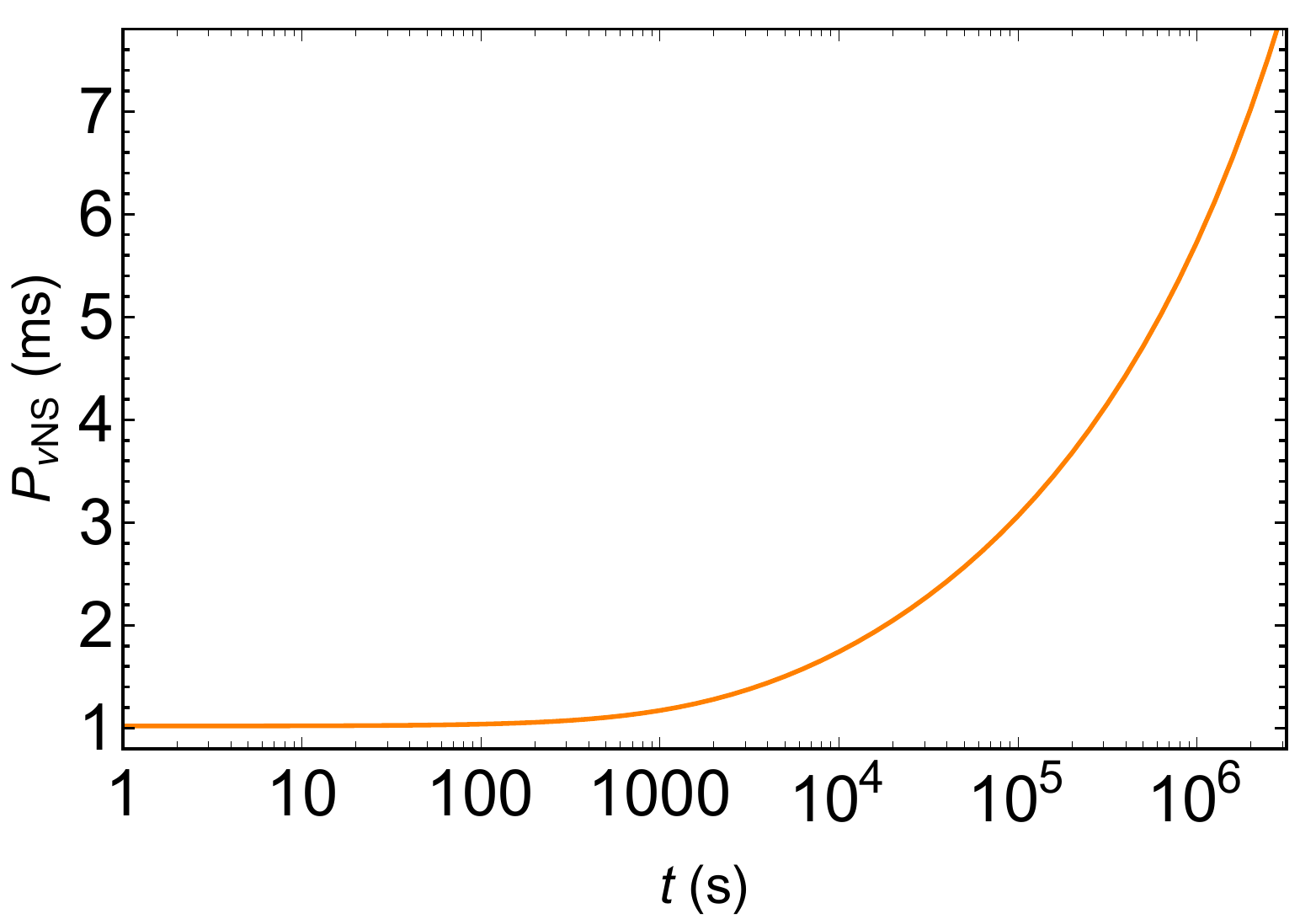}
\caption{Predicted time evolution of the rotation period of the $\nu$NS associated with the $\nu$NS rotational energy loss shown in Fig.~\ref{fig:lumthNS}.}
\label{fig:Pvst}
\end{figure}

\begin{table*}[!ht]
\centering
\caption{X-ray observational properties and inferred physical quantities of the $\nu$NS of the corresponding BdHN model that fits the data of GRB 180720B. Column 1: GRB name; column 2: identified BdHN type; column 3: the isotropic energy released ($E_{\rm iso}$) in gamma-rays; column 4: cosmological redshift ($z$); column 5: $\nu$NS rotation period ($P_{\nu \rm NS}$), column 6: $\nu$NS rotational energy ($E_{\rm rot}$); columns 7 and 8: strength of the dipole ($B_{\rm dip}$) and quadrupole ($B_{\rm quad}$) magnetic field components of the $\nu$NS. The quadruple magnetic field component is given in a range that the upper limit is three times than the lower limit, this is brought by the freedom of inclination angles of the magnetic moment.  During the fitting, we consistently assume the NS mass of $1.4 M_\odot$ and the NS radius of $10^{6}$~cm. The fitted light-curve is shown in Fig.~\ref{fig:lumthNS}. We refer the reader to \citet{2019ApJ...874...39W} for details about this model.}
\label{tab:ns_parameters}
\small
\begin{tabular}{@{}cccccccc@{}}
\hline
GRB     & Type    & Redshift & $E_{\rm iso}$  & $P_{\nu \rm NS}$  & $E_{\rm rot}$ & $B_{\rm dip}$ & $B_{\rm quad}$\\ 
        &         &          & (erg) & (ms) & (erg) & (G) & (G) \\
180720B & BdHN I  & 0.654     & $6.8\times10^{53}$& 0.5  & $4.03\times10^{52}$ & $4.22 \times 10^{13}$ & $1.0 \times 10^{13} \sim 2.11 \times 10^{15}$  \\
\hline
\end{tabular}
\end{table*}

Adopting fiducial values of mass, $M = 1.4\,M_{\odot}$, radius, $R = 10$~km, and moment of inertia, $I=10^{45}$~g~cm$^2$, Table~\ref{tab:ns_parameters} reports the values of $B_{\rm dip}$, $B_{\rm quad}$, and $P_{\nu \rm NS}$ obtained from the fitting procedure. It is evident from this analysis, see Fig.~\ref{fig:lumthNS}, that the $\nu$NS emission is not able to explain the emission of the ``Nousek-Zhang'' phase at early times at $t\lesssim 5 \times 10^2$~s. As it is shown in \citet{2018ApJ...869..101R, 2018ApJ...869..151R}, the emission in this phase is produced by the kinetic energy of a mildly-relativistic SN (\textit{supernova dominated region}), energized in part by the $\nu$NS. After $5\times 10^2$~s, as also shown by \citet{2018ApJ...869..101R,2020ApJ...893..148R}, the emission is mostly powered by the $\nu$NS (\textit{pulsar dominated region}).

\subsection{The BdHN theory and the nature of X-ray, GeV and H.E.S.S. radiations}\label{sec:bdhne}

The BdHN model has three components, the SN, the $\nu$NS and the BH. The X-ray afterglow luminosity presented in section~\ref{sec:x_ray} has allowed us to determine the spin and magnetic field of the $\nu$NS as presented in section~\ref{sec:nuns}. The temporal power-law behavior of the luminosity determines how the spin of the $\nu$NS evolves.

Likewise, the GeV luminosity presented in section~\ref{sec:G_ray} allows us to determine the mass and spin of the BH as presented in section~\ref{sec:inner}. The temporal power-law behavior of the luminosity determines how the spin of the BH evolves.

The best fits of the late X-ray and GeV luminosity power-laws of BdHN I GRB 180720B are shown in Fig.~\ref{fig:3pdf}. The X-ray afterglow luminosity observed by \textit{Swift}-XRT fades with time following a power-law behavior with amplitude of $(2.5\pm 0.4)\times 10^{53}$~erg~s$^{-1}$ and a power-law index $\alpha_X=1.43\pm 0.07$. It has been shown in \citet{2018ApJ...852...53R, 2018ApJ...869..151R} that, till $\sim 10^3$~s, the gamma/X-ray afterglow is mainly produced by the SN activities mainly powered by the mildly-relativistic SN kinetic energy. After $\sim 10^3$~s, the role of the $\nu$NS becomes prominent \citep{2018ApJ...869..101R}. 

The light-curve of Fermi-LAT fades with time following a power-law with an amplitude $A_{\rm GeV} = (4.6\pm 2.9)\times 10^{53}$~erg~s$^{-1}$ and index $\alpha_{\rm GeV}=1.94\pm 0.04$. This decaying behavior is clearly different from the one observed by Swift-XRT. This difference points to a different source for these radiations. As it can be seen from Fig.~\ref{fig:3pdf}, the extrapolation of the GeV luminosity does not coincide with the luminosity obtained from H.E.S.S. data. Instead, the H.E.S.S. luminosity lies on the power-law of the X-ray luminosity. This calls the attention that the nature of the radiation observed by H.E.S.S. might depend on the $\nu$NS activities. In the next sections, we probe the possibility that this radiation can be caused by the glitch events and injected relativistic electrons into the magnetosphere of the $\nu$NS during its slowing down.  

Also, as pointed out by \citet{2020A&A...636A..55R}, the LAT data and their temporal decay are insufficient to unveil the nature of the emission component detected by H.E.S.S.. Therefore, the radiations observed by LAT and H.E.S.S, overwhelmingly, are not originated from the same source.

\section{Glitches in Pulsar}\label{sec:glitch}

\begin{figure}
    \centering
    \includegraphics[width=\hsize,clip,angle=0]{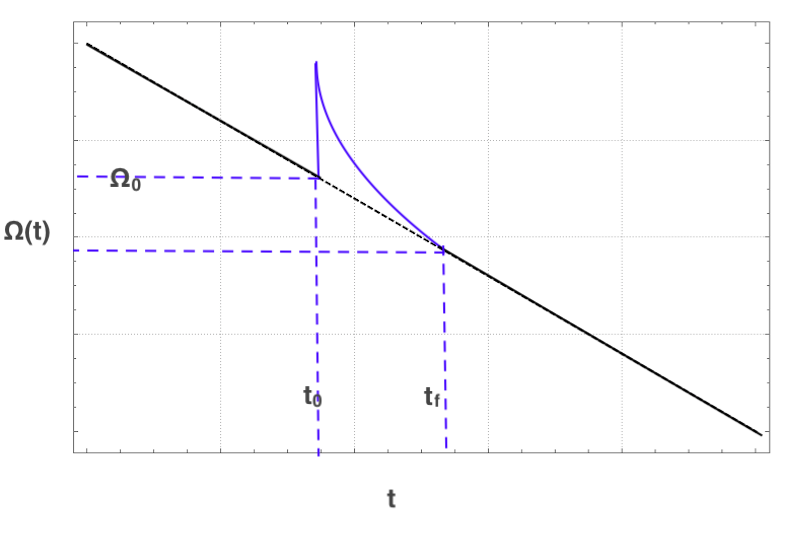}
    \caption{Qualitative spindown evolution of a pulsar experiencing a glitch with a full recovery of the pulsar pre-glitch spindown behavior.}
    \label{fig:glitch}
\end{figure}

We recall that glitches are sudden spin-up events observed in pulsars, as qualitatively shown in Fig.~\ref{fig:glitch}. First, let us consider the pulsar properties during the time interval from $t_0 = t_g$ and $t_f = t_g + \Delta t_{\rm rec}$, in absence of a glitch, which $\Delta t_{\rm rec}$ is the timescale of glitch recovery.

The pulsar would have lost an amount of rotational energy \citep{2012PASJ...64...56M}:
\begin{equation}
    \Delta T = \int_{t_0}^{t_f} |\dot{T}| dt,
\end{equation}
where $|\dot{T}|$ is the rotational energy loss of the pulsar:
\begin{equation}\label{eq:Erot}
    |\dot{T}| = I \Omega |\dot{\Omega}|,
\end{equation}
where the moment of inertia $I$ has been assumed to be nearly constant during the time interval $\Delta t_{\rm rec}$. The pulsar spindown timescale at $t_0$, can be estimated as
\begin{equation}\label{eq:tausd}
    \tau_{\rm sd,0} = \frac{\Omega_0}{|\dot{\Omega}_0|},
\end{equation}
which, for typical pulsar parameters, is much longer than the glitch recovery timescale, $\Delta t_{\rm rec}$. The rotational energy loss given by Eq.~(\ref{eq:Erot}) can then be approximated with good accuracy (of order $\Delta t_{\rm rec}/\tau_{\rm sd,0}$) by:
\begin{equation}\label{eq:DeltaE}
    \Delta T\approx |\dot{T}_0| \Delta t_{\rm rec} = I_0 \Omega_0 |\dot{\Omega}_0| \Delta t_{\rm rec},
\end{equation}
where the subscript `0' stands for pre-glitch values.

The angular velocity at $t_f$, within the same approximation, can be written as 
\begin{equation}\label{eq:Omf}
    \Omega_f = \Omega(t_g + \Delta t_{\rm rec})\approx \Omega_0\left(1 - \frac{\Delta t_{\rm rec}}{\tau_{\rm sd}}\right).
\end{equation}
It can be checked that the difference in rotational energy between $t_0$ and $t_f$, agrees with Eq.~(\ref{eq:DeltaE}):
\begin{equation}
     \Delta T = \frac{1}{2} I|\Omega_f^2 - \Omega_0^2|\approx I \Omega_0 |\dot{\Omega}_0| \Delta t_{\rm rec},
\end{equation}
where we have used Eqs.~(\ref{eq:tausd}) and (\ref{eq:Omf}).

Now, let us assume that the pulsar experiences at $t_0$ a glitch such as the one shown in Fig.~\ref{fig:glitch}. So, at the time $t_f$, the pulsar fully recovers (i.e. match) the spindown behavior of the pulsar expected from the pre-glitch stage. In such a case, it is clear that energy conservation implies that the energy released in the glitch and its recovery must be equal to the energy loss (\ref{eq:DeltaE}):
\begin{equation}\label{eq:DeltaEg}
    \Delta E_g = \Delta T = \dot{T}_0\Delta t_{\rm rec} = I \Omega_0 |\dot{\Omega}_0| \Delta t_{\rm rec}.
\end{equation}
Equation~(\ref{eq:DeltaEg}) gives crucial information of how the glitch energetics relates to the pulsar spindown behavior and the recovery time. Therefore, $\Delta E_g$ is the maximum energy that can be released in a full recovery glitch event, so the maximum observable energy (i.e. radiated off in electromagnetic waves), which becomes crucial for the analysis of the \emph{glitch-burst} connection. 

All the above equations show, in particular, that in young pulsars the total energy of the glitch can increase sharply in the early part of afterglow. It is important to underline that the $\Delta E_{g, \rm rad}$ can occur in a vast range of energies, ranging from the traditional X-ray all the way up to the TeV radiation. The case of the first glitch in GRB 180720B occurs, where the energy has reached H.E.S.S. radiation band with isotropic energy of $\Delta E_{g, \rm rad}=2.4 \times 10^{50}$~erg during the characteristic $\Delta t_{\rm rec}$ of $10^3$~s. 

This shows that the energetics of H.E.S.S. radiation can be indeed explained by a glitch event produced by the sudden spinup of the $\nu$NS. We advance the possibility that the mechanism of producing such a high-energy emission might be similar to the one leading to the late X-ray afterglow emission which is originated from the synchrotron emission of electrons accelerated in an expanding magnetized HN ejecta, as presented in \citet{2018ApJ...869..101R, 2019ApJ...874...39W, 2020ApJ...893..148R}. The only difference can be the number and relativistic nature of the electrons which, during the sudden spinup of the $\nu$NS, accelerate inside the magnetized HN and produce the high-energy emission. This differs from traditional works relating the presence of glitches to changes in the moment of inertia related to superconductivity or other turbulent phenomena \citep{1970Natur.227..791G,1972NPhS..235...43P,1975Natur.256...25A}. The precise details of the operation of such a mechanism deserve to be studied in forthcoming analyses to be presented elsewhere.

\section{Conclusion}\label{sec:conclusion}

GRB 180720B is complementary to the case of GRB 130427A \citep{2019ApJ...886...82R} and GRB 190114C \citep{2019arXiv190404162R, 2019arXiv191107552M} and offers a qualitative and quantitative confirmation of the expected BdHN I nature of these three sources. In addition to these confirmations, each of these sources has revealed the existence of a new physical process whose discovery extends the domain of the fundamental physical laws of physics and more specific of relativistic astrophysics.

Thanks to the comparison of the observations of these sources with their theoretical astrophysical description, it has been confirmed in the recent years that the progenitor of a BdHN I is a binary system composed of a CO$_{\rm core}$ and a companion NS; see e.g. \citet{2016ApJ...832..136R, 2016ApJ...833..107B, 2018ApJ...852...53R, 2019ApJ...871...14B, 2019ApJ...874...39W,2019ApJ...886...82R, 2020ApJ...893..148R}, and Ruffini, et al. (submitted). The triggering of the GRB is manifested by the observation of the \textit{SN-rise}, lasting a fraction of a second, and which has been observed in all these GRBs with a clear spectral signature of a BB component. It has been also verified that the binary periods of all three sources is of the order of $5$~min, which has confirmed our standard description of the hypercritical accretion of the SN ejecta onto the companion NS, leading to the BH formation. The latter is heralded by the onset of the GeV emission which has been observed in these sources. Still in these three sources, the emission of the \textit{cavity} created by the BH formation, with its theoretically understood featureless spectrum \citep{2019ApJ...883..191R}, has been observed; see \citet{Li2019}, Ruffini, et al. (submitted), and Figs.~\ref{fig:lc_180720B} and \ref{fig:3episodes}.

One of the three main discoveries of the BdHN I has been the UPE phase observed both in GRB 190114C \citep{2019arXiv190404162R} and in GRB 180720B; see Fig.~\ref{fig:lc_180720B}. In both cases, the self-similar structure has been discovered; see Fig.~\ref{fig:kT-Luminosity-evolution} and Table~\ref{tab:180720B} in appendix~\ref{sec:time-resolved}. In the case of GRB 130427A, the data of the UPE phase could not be analyzed since they were affected by a strong pile-up effect \citep{2014Sci...343...42A, 2015ApJ...798...10R}. 

The pile-up in GRB 130427A turns out to be a fortunate coincidence: we addressed the analysis of the GeV radiation following the UPE phase and this has led to the first major new discovery in the BdHN I astrophysics: the \textit{inner engine} with the power-law emission of the GeV luminosity \citep{2019ApJ...886...82R} which has been also confirmed in GRB 190114C \citep{2019arXiv191107552M} and in GRB 180720B (see Fig.~\ref{fig:3pdf}). The understanding of the \textit{inner engine} has been essential in identifying in the rotational energy of the BH as the source of the power-law GeV luminosity, which has led to the determination, for the first time, of the BH mass of $M=2.32 M_\odot$, of the spin parameter of $\alpha = 0.27$ and of the magnitude of surrounding magnetic field of $\approx 6\times 10^{10}$~G; see section~\ref{sec:inner} (see also section~$6$ of \citealp{2019ApJ...886...82R} in the case of GRB 130427A). The canonical value of the observed masses and spin of the BH in all three sources can certainly be considered a great conceptual success.

The X-ray afterglow emission, previously discovered to originate in the synchrotron emission of the expanding SN, moving in the magnetic field of the $\nu$NS, and energized by the $\nu$NS emission powered by hypercritical accretion of SN matter fallback and magnetic braking \citep{2018ApJ...869..101R,2020ApJ...893..148R}, had signed a turning point in the understanding of the BdHN I nature. Indeed, in all the above sources, these results have been led to the determination of: a) the spin of the $\nu$NS at birth, $P_{\rm \nu NS,0}\sim 1$~ms, b) the magnetic field multipolar structure with a dipole strength of $B\sim 10^{13}$~G, and c) the explanation of the X-ray luminosity, monotonically decreasing with time as a power-law, observed in all three systems; see Figs.~\ref{fig:lumthNS} and \ref{fig:3pdf}.

The outstanding contributions of GRB 180720B to fundamental GRB science have come from
\begin{itemize}
    \item 
    the presence of TeV emission observed by H.E.S.S in $100$--$440$~GeV after $10.1$~h, with an isotropic energy of $E^{\rm iso}_{\rm HESS}= (2.4\pm 1.8) \times 10^{50}$~erg and duration of $\sim 1.2$~h; see sections~\ref{sec:aftert90} and \ref{sec:BH-NS-Hess};
    \item 
    the observation of the GeV emission occurring at an intermediate angle between the equatorial plane of the binary and the BH rotation axis;
    \item 
    the corresponding modulation of the UPE emission and the appearance of the soft and hard X-ray flares; see Figs.~\ref{fig:1pdf}, \ref{fig:2pdf} and \ref{fig:3pdf}.
\end{itemize}

We have here explained in Fig.~\ref{fig:3pdf} and in section~\ref{sec:glitch} the possible connection of the sub-TeV emission with a pulsar glitch, a possibility which is intrinsic in our $\nu$NS interpretation of the X-ray afterglow. The amplitude of the glitches is strongly correlated to the period of the pulsar. By duly extrapolating this behavior to GRB 190114C, we can also explain the TeV emission observed by MAGIC \citep{2019GCN.23701....1M}.

We have decided to address the nature of the UPE in the simplest possible case: the one of GRB 190114C where the GRB is seen ``from the top'' of the binary progenitor (Moradi, et al., to be submitted). This is the simplest case to have the first basic  understanding of the vacuum polarization processes \citep{1975PhRvL..35..463D,RSWX2,RSWX} observed in the UPE phase, opening an additional new physics scenario; \citealp[see also][]{2010PhR...487....1R}.

The common outcome of these three sources which brings the greatest new message for the BH physics is:
\begin{enumerate}
    \item 
    in the astrophysical reality a Kerr BH is always embedded in a magnetic field likely aligned with its rotation axis;
    \item 
    the BH is not in vacuum but embedded in a very-low- density plasma;
    \item 
    the BH rotational energy extraction does allow the GeV radiation to endure, in principle, for a cosmological time and the fate of a BH is entangled with the final evolution of the Universe and, until then, it is alive (see, e.g., P.~$545$ and Part $\rm IV$: ``Energetics of Black Holes'' of ~\citealp{DeWitt:1973uma} by R. Ruffini). Therefore, we conclude that the issue of observing a singularity inside a Kerr BH has no astrophysical foundation; see also \href{https://www.youtube.com/watch?v=ztWc_pj-4EA}{``The 2020 Oskar Klein Memorial Lecture: \textit{Kerr Black Holes have no Singularities} by Roy P. Kerr''}.
\end{enumerate}


\appendix

\section{Time-resolved spectral analysis of the UPE phase of GRB 180720B}\label{sec:time-resolved}

\begin{figure*}
\includegraphics[angle=0, scale=0.47]{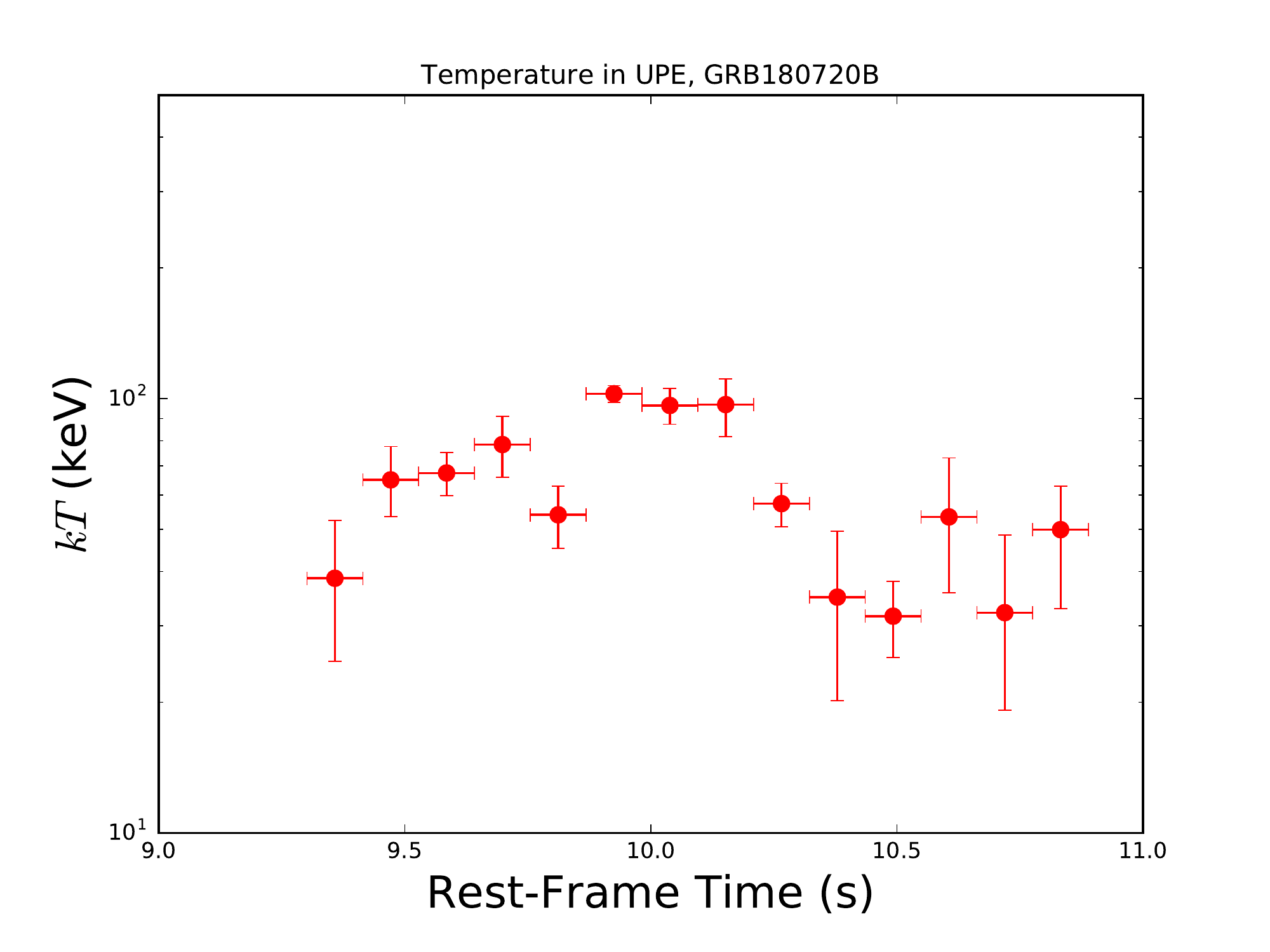}
\includegraphics[angle=0, scale=0.47]{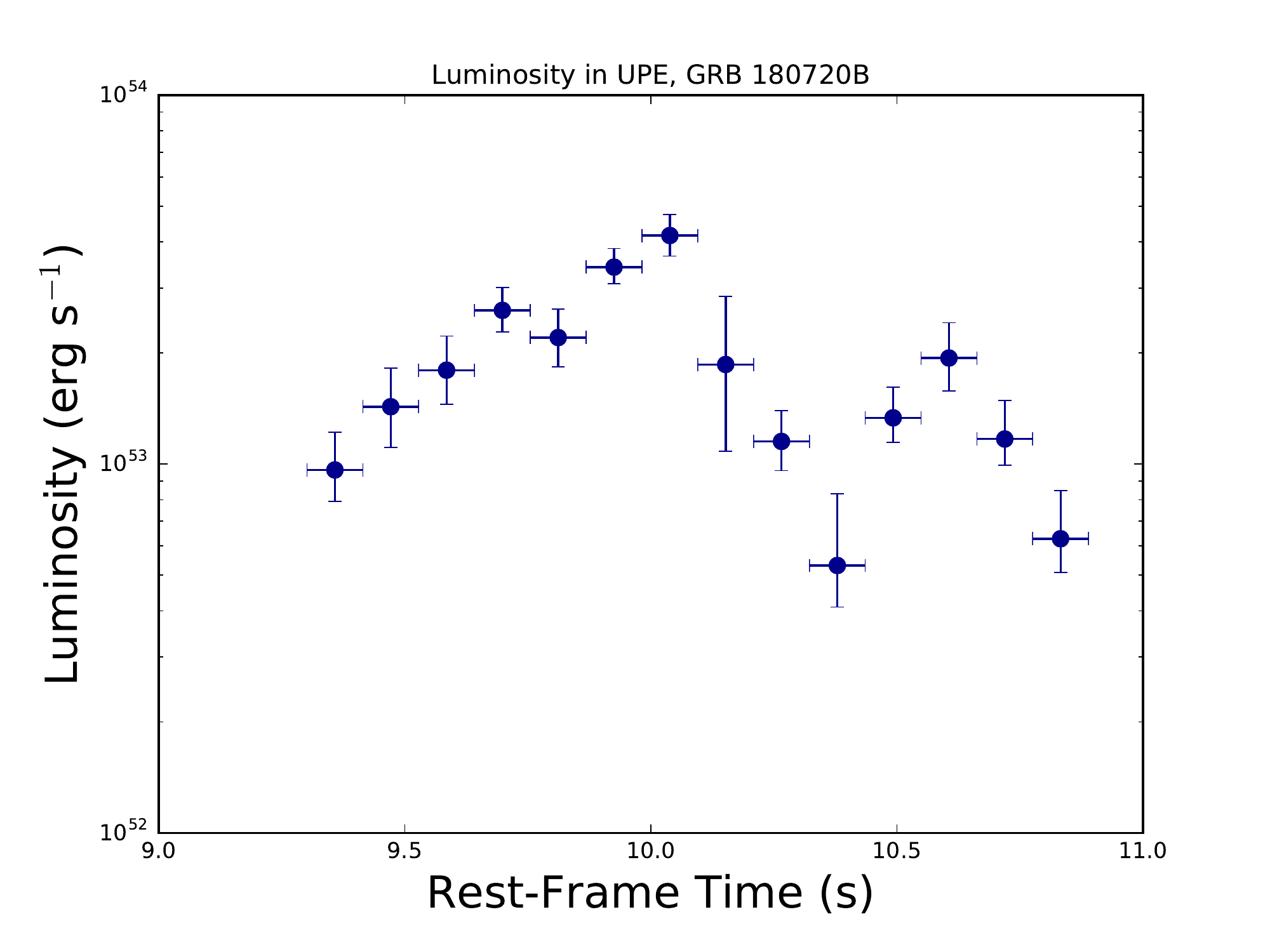}
\caption{The temperature ($kT$) and the luminosity evolution during the UPE phase of GRB 1807820B obtained from time-resolved spectral analysis of the Fermi-GBM data.}\label{fig:kT-Luminosity-evolution}
\end{figure*}

The high signal-to-noise (S/N) ratio of the data of Episode $2$ has allowed us to perform a refined spectral analysis in the $[9.07$--$10.89]$ time interval in $5$ iterations on decreasing time bins, keeping still the reliable statistical significance: in each time bin a CPL+BB spectrum is found as the best fit. The time intervals both in rest-frame and observer frame, the significance ($S$) for each time interval, the power-law index, cut-off energy, temperature, $\Delta$DIC, BB flux, total flux, the BB to total flux ratio, $F_{\rm BB}/F_{\rm tot}$ and finally the isotropic energy of entire this time bin and its sub-intervals are shown in Table~\ref{tab:180720B}. The evolution of the temperature and the luminosity of UPE phase, as obtained by the time-resolved spectral analysis are shown in Fig.~\ref{fig:kT-Luminosity-evolution}.

\begin{deluxetable*}{ccccccccccc}
\tabletypesize{\scriptsize}
\tablecaption{Results of the time-resolved spectral fits of GRB 180720B (CPL+BB model) from the $t_{\rm rf}=9.07$~s to $t_{\rm rf}=10.89$~s. This table reports: the time intervals both in rest-frame and observer frame, the significance ($S$) for each time interval, the power-law index, cut-off energy, temperature, $\Delta$DIC, BB flux, total flux, the BB to total flux ratio, $F_{\rm BB}/F_{\rm tot}$ and finally the isotropic energy. To select the best model from two different given models, we adopt the deviance information criterion (DIC), defined as DIC=-2log[$p$(data$\mid\hat{\theta}$)]+2$p_{\rm DIC}$, where $\hat{\theta}$ is the posterior mean of the parameters, and $p_{\rm DIC}$ is the effective number of parameters. The preferred model is the model with the lowest DIC score. Here we define $\Delta$DIC=(CPL+BB)-CPL, if $\Delta$DIC is negative, indicating the CPL+BB is better. After comparing the DIC, we find the CPL+BB model is the preferred model than the CPL and other model. The $\Delta$DIC scores are reported in column 6. The redshift z=0.653 \label{tab:180720B}}
\tablehead{
\colhead{$t_{1}$$\sim$$t_{2}$}
&\colhead{$t_{1}$$\sim$$t_{2}$}
&\colhead{$S$}
&\colhead{$\alpha$}
&\colhead{$E_{\rm c}$}
&\colhead{$kT$}
&\colhead{$\Delta$DIC}
&\colhead{$F_{\rm BB}$}
&\colhead{$F_{\rm tot}$}
&\colhead{$F_{\rm ratio}$}
&\colhead{$E_{\rm tot}$}\\
\hline
(s)&(s)&&&(keV)&(keV)&&(10$^{-6}$)&(10$^{-6}$)&&(10$^{52}$)\\
Obs&Rest&&&&&&(erg~cm$^{-2}$~s$^{-1}$)&(erg~cm$^{-2}$~s$^{-1}$)&&erg
}
\colnumbers
\startdata
15.00$\sim$18.00&9.07$\sim$10.89&274.60&-1.06$^{+0.01}_{-0.01}$&1502.5$^{+88.6}_{-87.5}$&39.8$^{+1.6}_{-1.6}$&-226.4&1.99$^{+0.43}_{-0.34}$&45.55$^{+3.11}_{-2.70}$&0.04$^{+0.01}_{-0.01}$&16.0$^{+1.1}_{-0.952}$\\
\hline
15.00$\sim$16.50&9.07$\sim$9.98&190.63&-1.04$^{+0.01}_{-0.01}$&1750.5$^{+112.7}_{-111.1}$&40.5$^{+2.0}_{-2.0}$&-176.6&2.08$^{+0.58}_{-0.46}$&48.03$^{+3.28}_{-3.09}$&0.04$^{+0.01}_{-0.01}$&8.46$^{+0.577}_{-0.543}$\\
16.50$\sim$18.00&9.98$\sim$10.89&215.76&-1.05$^{+0.02}_{-0.02}$&1151.3$^{+117.3}_{-119.6}$&37.1$^{+2.8}_{-2.8}$&-78.7&1.63$^{+0.69}_{-0.54}$&41.83$^{+4.61}_{-4.04}$&0.04$^{+0.02}_{-0.01}$&7.37$^{+0.812}_{-0.712}$\\
\hline
15.00$\sim$15.75&9.07$\sim$9.53&105.93&-1.07$^{+0.03}_{-0.03}$&1198.0$^{+211.1}_{-217.8}$&31.4$^{+3.3}_{-3.3}$&-41.5&0.94$^{+0.70}_{-0.42}$&23.84$^{+4.65}_{-3.86}$&0.04$^{+0.03}_{-0.02}$&2.1$^{+0.41}_{-0.34}$\\
15.75$\sim$16.50&9.53$\sim$9.98&168.59&-0.92$^{+0.02}_{-0.02}$&1028.0$^{+74.9}_{-73.9}$&74.8$^{+20.8}_{-25.7}$&-15.4&0.14$^{+0.37}_{-0.13}$&58.57$^{+5.42}_{-4.80}$&0.0$^{+0.01}_{-0.0}$&5.16$^{+0.478}_{-0.423}$\\
16.50$\sim$17.25&9.98$\sim$10.44&155.67&-1.15$^{+0.02}_{-0.02}$&2382.3$^{+217.5}_{-221.3}$&45.3$^{+2.7}_{-2.7}$&-125.6&2.85$^{+1.00}_{-0.76}$&53.96$^{+4.55}_{-4.28}$&0.05$^{+0.02}_{-0.01}$&4.75$^{+0.401}_{-0.377}$\\
17.25$\sim$18.00&10.44$\sim$10.89&159.05&-0.93$^{+0.02}_{-0.02}$&684.7$^{+49.7}_{-49.2}$&23.9$^{+3.8}_{-4.0}$&-30.8&0.63$^{+0.93}_{-0.37}$&35.74$^{+3.28}_{-3.21}$&0.02$^{+0.03}_{-0.01}$&3.15$^{+0.289}_{-0.283}$\\
\hline
15.00$\sim$15.38&9.07$\sim$9.30&69.11&-1.06$^{+0.07}_{-0.08}$&711.2$^{+209.5}_{-215.5}$&28.9$^{+5.7}_{-5.6}$&-30.2&0.78$^{+1.14}_{-0.55}$&14.27$^{+6.80}_{-3.54}$&0.05$^{+0.08}_{-0.04}$&0.628$^{+0.299}_{-0.156}$\\
15.38$\sim$15.75&9.30$\sim$9.53&83.03&-1.01$^{+0.03}_{-0.03}$&1319.4$^{+210.9}_{-208.7}$&31.0$^{+5.2}_{-5.2}$&-28.9&0.83$^{+1.14}_{-0.48}$&32.18$^{+6.45}_{-5.45}$&0.03$^{+0.04}_{-0.02}$&1.42$^{+0.284}_{-0.24}$\\
15.75$\sim$16.12&9.53$\sim$9.75&109.59&-1.02$^{+0.02}_{-0.02}$&1967.9$^{+193.8}_{-194.9}$&43.6$^{+4.0}_{-4.0}$&-72.6&2.63$^{+1.51}_{-0.96}$&62.61$^{+6.83}_{-6.58}$&0.04$^{+0.02}_{-0.02}$&2.76$^{+0.301}_{-0.29}$\\
16.12$\sim$16.50&9.75$\sim$9.98&133.10&-1.01$^{+0.02}_{-0.02}$&1919.4$^{+162.1}_{-168.5}$&47.9$^{+3.5}_{-3.5}$&-107.5&4.31$^{+1.60}_{-1.38}$&82.08$^{+8.46}_{-7.17}$&0.05$^{+0.02}_{-0.02}$&3.61$^{+0.372}_{-0.316}$\\
16.50$\sim$16.88&9.98$\sim$10.21&133.12&-1.09$^{+0.02}_{-0.02}$&2574.3$^{+264.0}_{-267.2}$&55.7$^{+3.8}_{-3.7}$&-117.9&5.16$^{+2.03}_{-1.44}$&83.97$^{+8.79}_{-7.60}$&0.06$^{+0.02}_{-0.02}$&3.7$^{+0.387}_{-0.335}$\\
16.88$\sim$17.25&10.21$\sim$10.44&89.16&-1.24$^{+0.05}_{-0.05}$&1537.9$^{+522.7}_{-558.0}$&31.9$^{+3.4}_{-3.4}$&-27.8&1.38$^{+0.94}_{-0.57}$&24.25$^{+7.37}_{-6.29}$&0.06$^{+0.04}_{-0.03}$&1.07$^{+0.325}_{-0.277}$\\
17.25$\sim$17.62&10.44$\sim$10.66&125.76&-0.86$^{+0.03}_{-0.03}$&696.1$^{+59.2}_{-57.7}$&22.5$^{+3.8}_{-3.7}$&-27.3&0.83$^{+1.39}_{-0.48}$&45.89$^{+5.21}_{-4.69}$&0.02$^{+0.03}_{-0.01}$&2.02$^{+0.23}_{-0.206}$\\
17.62$\sim$18.00&10.66$\sim$10.89&102.97&-1.02$^{+0.04}_{-0.04}$&622.4$^{+77.4}_{-80.6}$&25.7$^{+8.4}_{-9.5}$&-25.5&0.39$^{+1.32}_{-0.34}$&25.51$^{+4.95}_{-3.40}$&0.02$^{+0.05}_{-0.01}$&1.12$^{+0.218}_{-0.15}$\\
\hline
15.00$\sim$15.19&9.07$\sim$9.19&51.57&-1.01$^{+0.14}_{-0.15}$&805.3$^{+449.1}_{-380.0}$&33.0$^{+13.1}_{-18.2}$&-288.8&0.80$^{+5.09}_{-0.77}$&19.23$^{+23.49}_{-7.86}$&0.04$^{+0.27}_{-0.04}$&0.423$^{+0.517}_{-0.173}$\\
15.19$\sim$15.38&9.19$\sim$9.30&42.03&-1.19$^{+0.09}_{-0.09}$&1201.3$^{+667.6}_{-595.4}$&27.5$^{+4.3}_{-4.2}$&-27.1&0.97$^{+1.06}_{-0.55}$&12.89$^{+8.98}_{-4.04}$&0.08$^{+0.1}_{-0.05}$&0.284$^{+0.198}_{-0.0889}$\\
15.38$\sim$15.56&9.30$\sim$9.41&53.84&-1.00$^{+0.04}_{-0.04}$&1158.5$^{+201.4}_{-200.2}$&23.4$^{+8.3}_{-8.3}$&-27.1&0.29$^{+1.66}_{-0.26}$&27.59$^{+7.34}_{-4.93}$&0.01$^{+0.06}_{-0.01}$&0.608$^{+0.162}_{-0.109}$\\
15.56$\sim$15.75&9.41$\sim$9.53&63.61&-1.06$^{+0.05}_{-0.05}$&1839.8$^{+434.0}_{-420.6}$&39.4$^{+7.6}_{-7.0}$&-32.2&1.74$^{+2.60}_{-1.11}$&40.95$^{+11.15}_{-9.24}$&0.04$^{+0.06}_{-0.03}$&0.902$^{+0.246}_{-0.203}$\\
15.75$\sim$15.94&9.53$\sim$9.64&72.54&-1.04$^{+0.04}_{-0.04}$&1896.8$^{+350.9}_{-351.5}$&40.8$^{+4.7}_{-4.7}$&-30.3&2.78$^{+2.13}_{-1.19}$&51.44$^{+12.23}_{-9.91}$&0.05$^{+0.04}_{-0.03}$&1.13$^{+0.269}_{-0.218}$\\
15.94$\sim$16.12&9.64$\sim$9.75&83.99&-0.99$^{+0.03}_{-0.03}$&1950.2$^{+231.8}_{-232.1}$&47.5$^{+7.6}_{-7.6}$&-34.3&2.34$^{+3.12}_{-1.29}$&74.72$^{+11.53}_{-9.35}$&0.03$^{+0.04}_{-0.02}$&1.65$^{+0.254}_{-0.206}$\\
16.12$\sim$16.31&9.75$\sim$9.87&85.09&-0.95$^{+0.04}_{-0.04}$&1379.2$^{+207.4}_{-203.8}$&32.7$^{+5.4}_{-5.3}$&-39.2&1.84$^{+2.29}_{-1.02}$&63.06$^{+12.29}_{-10.56}$&0.03$^{+0.04}_{-0.02}$&1.39$^{+0.271}_{-0.233}$\\
16.31$\sim$16.50&9.87$\sim$9.98&104.94&-1.05$^{+0.02}_{-0.02}$&2304.7$^{+260.1}_{-261.8}$&62.1$^{+2.8}_{-2.8}$&-85.4&6.72$^{+1.63}_{-1.29}$&97.87$^{+12.08}_{-9.75}$&0.07$^{+0.02}_{-0.01}$&2.15$^{+0.266}_{-0.215}$\\
16.50$\sim$16.69&9.98$\sim$10.10&107.18&-1.04$^{+0.03}_{-0.03}$&2737.1$^{+346.9}_{-340.9}$&58.4$^{+5.6}_{-5.6}$&-86.1&6.57$^{+3.89}_{-2.56}$&119.20$^{+16.65}_{-14.38}$&0.06$^{+0.03}_{-0.02}$&2.62$^{+0.367}_{-0.317}$\\
16.69$\sim$16.88&10.10$\sim$10.21&82.58&-1.13$^{+0.13}_{-0.08}$&1910.0$^{+709.1}_{-1074.0}$&58.6$^{+8.6}_{-9.2}$&-86.9&3.67$^{+4.06}_{-3.43}$&53.29$^{+28.29}_{-22.24}$&0.07$^{+0.08}_{-0.07}$&1.17$^{+0.623}_{-0.49}$\\
16.88$\sim$17.06&10.21$\sim$10.32&64.96&-1.24$^{+0.03}_{-0.03}$&2412.4$^{+580.9}_{-576.0}$&34.7$^{+4.0}_{-4.0}$&-28.1&1.52$^{+1.46}_{-0.72}$&32.97$^{+6.96}_{-5.49}$&0.05$^{+0.05}_{-0.02}$&0.726$^{+0.153}_{-0.121}$\\
17.06$\sim$17.25&10.32$\sim$10.44&62.39&-1.06$^{+0.08}_{-0.08}$&480.3$^{+112.6}_{-114.6}$&21.1$^{+8.8}_{-8.9}$&-125.2&0.39$^{+3.01}_{-0.35}$&15.20$^{+8.60}_{-3.47}$&0.03$^{+0.2}_{-0.02}$&0.335$^{+0.189}_{-0.0764}$\\
17.25$\sim$17.44&10.44$\sim$10.55&81.92&-0.89$^{+0.05}_{-0.05}$&720.6$^{+93.9}_{-92.3}$&19.1$^{+3.9}_{-3.8}$&-23.5&0.82$^{+1.62}_{-0.55}$&38.20$^{+8.11}_{-5.42}$&0.02$^{+0.04}_{-0.01}$&0.841$^{+0.179}_{-0.119}$\\
17.44$\sim$17.62&10.55$\sim$10.66&97.68&-0.84$^{+0.05}_{-0.05}$&713.4$^{+96.8}_{-97.0}$&32.3$^{+11.9}_{-10.7}$&-38.1&1.05$^{+5.66}_{-0.87}$&55.49$^{+13.70}_{-10.34}$&0.02$^{+0.1}_{-0.02}$&1.22$^{+0.302}_{-0.228}$\\
17.62$\sim$17.81&10.66$\sim$10.78&82.29&-0.95$^{+0.05}_{-0.05}$&628.7$^{+86.6}_{-86.2}$&19.5$^{+9.9}_{-7.8}$&-66.8&0.33$^{+4.15}_{-0.30}$&33.47$^{+9.11}_{-5.06}$&0.01$^{+0.12}_{-0.01}$&0.737$^{+0.201}_{-0.111}$\\
17.81$\sim$18.00&10.78$\sim$10.89&64.36&-1.08$^{+0.06}_{-0.06}$&565.9$^{+123.9}_{-118.5}$&30.2$^{+7.8}_{-10.3}$&-15.3&0.36$^{+1.63}_{-0.33}$&17.96$^{+6.32}_{-3.42}$&0.02$^{+0.09}_{-0.02}$&0.395$^{+0.139}_{-0.0752}$\\
\enddata
\end{deluxetable*}

\end{document}